\documentclass[fleqn,usenatbib]{mnras}
\usepackage{newtxtext,newtxmath}
\usepackage[T1]{fontenc}
\DeclareRobustCommand{\VAN}[3]{#2}
\let\VANthebibliography\thebibliography
\def\thebibliography{\DeclareRobustCommand{\VAN}[3]{##3}\VANthebibliography}
\usepackage{graphicx}
\usepackage{subfigure}
\usepackage{amsmath}
\usepackage{makecell}

\title[Bicoherence results in MAXI J1535$-$571]{The bicoherence study of quasi-periodic oscillations in MAXI J1535$-$571}
\author[Zhu et al.]{
Ziyuan Zhu,
Xiao Chen\thanks{E-mail: xiao.chen@whu.edu.cn},
Wei Wang \thanks{E-mail: wangwei2017@whu.edu.cn}
\\
Department of Astronomy, School of Physics and Technology, Wuhan University, Wuhan 430072, China\\
}
\date{}
\pubyear{2024}

\begin{document}
\label{firstpage}
\pagerange{\pageref{firstpage}--\pageref{lastpage}}
\maketitle

\begin{abstract}
Bicoherence is a way to measure the phase coupling of triplets of Fourier frequencies. We use this method to analyze quasi-periodic oscillations (QPOs) in the black hole X-ray binary MAXI J1535$-$571 during its 2017 September-October outburst. The bicoherence provides an interesting new diagnostic to uncover QPO behaviour and the relationships between QPO harmonics and broadband noise. The bicoherence pattern of type-C QPOs starts as a 'web' pattern and changes to a 'hypotenuse' pattern after the presence of type-B QPOs, indicating that MAXI J1535$-$571 is a low-inclination source. The intensity of bicoherence also exhibits variations across different energy bands. We try to explain the bicoherence results in the scenario of a dual-corona geometry.
\end{abstract}


\section{Introduction}

The black hole X-ray binaries (BHXRBs) generally stay in quiescence for most of the lifetime. In a typical outburst, a BHXRB as an X-ray transient follows a 'q' path in the hardness-intensity diagram (HID; \citealp{Homan_2001, 10.1111/j.1365-2966.2004.08384.x}): it starts from the low hard state (LHS), goes through the hard intermediate state (HIMS) and soft intermediate state (SIMS), reaches the high soft state (HSS), and then returns to the LHS. Hysteresis phenomenon is sometimes exhibited in the HID \citep{Maccarone.2003.06040.x}.

Multiple models are proposed to interpret the origin of the LFQPOs, generally classified into intrinsic and geometric variability models. Intrinsic models suggest that LFQPOs are generated from instabilities in properties such as pressure or accretion rate, while geometric models often consider Lense-Thirring precession as the origin of the LFQPOs \citep{Stella_1998, Ingram.2009.00693.x}.

The transition from the HIMS to SIMS in BHXRBs is always accompanied by a change from type-C QPOs to type-B QPOs \citep{Wijnands1999ApJ...526L..33W, Casella2005ApJ...629..403C}. A shift in the radio jet is also detected during this period. The steady, compact jet can be quenched and replaced by a bright transient jet, along with relativistic ejecta \citep{10.1111/j.1365-2966.2004.08384.x, Tetarenko2017MNRAS.469.3141T}. However, there have been only a limited number of transient jets directly resolved in BHXRBs so far \citep{Mirabel1994Natur.371...46M, Hjellming1995Natur.375..464H, Yang2010MNRAS.409L..64Y, Miller-Jones2019Natur.569..374M}. Thus, the physical origin of the jet remains to be determined, and the clear connection between the discrete, transient jet and the type-B QPO still needs to be found \citep {Fender2009MNRAS.396.1370F, Miller-Jones2012MNRAS.421..468M}.

The most common way to analyze QPOs is to fit the power density spectrum (PDS) with a multi-Lorentzian function \citep{Belloni2002ApJ...572..392B}. The light curves are generally divided into several segments with the same length of time for statistical accuracy. The final PDS of a light curve is the average of the PDS in each segment. Since different models may generate similar PDSs, it is important to explore other timing techniques that can resolve ambiguities between them. One such technique is the bispectrum, which is capable of measuring the phase coupling of three specific frequencies \citep{Maccarone2004.08615.x}. This kind of phase coupling has been observed in many sources, such as GX 339-4 and GRS 1915+105 \citep{Arur_2019, Maccarone_2011}.

MAXI J1535-571 is an X-ray transient discovered independently by \textit{MAXI}/GSC \citep{Negoro2017ATel10708....1N} and \textit{Swift}/BAT \citep{Kennea2017ATel10700....1K} on September 02, 2017. Subsequent observations on X-ray and other wavelengths have been conducted since then \citep{Nakahira2017ATel10729....1N, Kenneaalone2017ATel10731....1K, Palmer2017ATel10733....1P, Tetarenko2017ATel10745....1T, Shidatsu2017ATel10761....1S}. Several studies indicate that this source is a black hole with a high spin and a high inclination. Based on \textit{NICER} data, \cite{Miller2018ApJ...860L..28M} suggested the spin to be $0.994\pm0.002$ and the inclination angle to be $67.4^\circ\pm0.8^\circ$. \cite{Xu2018ApJ...852L..34X} derived similar results with a spin $>0.84$ and an inclination angle of $\sim60^\circ$.

Approximately one week after the discovery of MAXI 1535-571, this BHXRB entered HIMS \citep{Kenneaalone2017ATel10731....1K, Nakahira2017ATel10729....1N, Palmer2017ATel10733....1P, Shidatsu2017ATel10761....1S, Tetarenko2017ATel10745....1T} with type-C QPOs detected \citep{Gendreau2017ATel10768....1G, Mereminskiy2017ATel10734....1M, Huang2018ApJ...866..122H,Bhargava2019MNRAS.488..720B, Vincentelli2021MNRAS.503..614V}. Within one to two months, it shifted into the SIMS and soft state \citep{Shidatsu2017ATel11020....1S}, while showing type-B QPOs during the SIMS \citep{Stevens2018ApJ...865L..15S}. At the end of September 2017, type-C QPOs reappeared when the source transitioned back to the HIMS and LHS \citep{Rawat2023MNRAS.520..113R}.

In this paper, we use the bicoherence method combined with PDS to analyze the phase coupling between QPOs and broadband noises in the MAXI J1535$-$571, using data collected by the \textit{Neutron Star Interior Composition Explorer} (\textit{NICER}). Section \ref{data} describes the observation and the data analysis approach, detailing the derivation of PDS and the statistical approach utilized, namely the bicoherence method. In Section \ref{result}, we present the \textbf{results} of the bicoherence patterns and PDS in this source. Finally, we discuss the results in Section \ref{discuss} and make a summary in Section \ref{summ}.

\section{Observation and data analysis}
\label{data}

\begin{table*}
    \centering
    \caption{Observation log of MAXI J1535-571, including timing parameters. The columns are the observation number, the NICER ObsID, the start and end time of the observation, the source count rate in the 1-10 keV band, the hardness ratio  (HR), the QPO centroid frequency, and the exposure time.}
    \label{}
    \begin{tabular}{cccccccc}
    \hline
    Obs no. & ObsID & \makecell{Tstart\\(M.J.D)} & \makecell{Tstop\\(M.J.D)} & \makecell{Count rate\\(1-10 keV)} & \makecell{HR\\$\frac{(3-10 keV)}{(1-3 keV)}$} & \makecell{QPO frequency\\(Hz)} & \makecell{Exposure Time\\(s)}\\
    \hline
    1 & 1050360104 & 58008.454 & 58008.982 & 7121 & 0.604 & 2.56 $\pm$ 0.01 & 5296\\
    2 & 1050360105 & 58008.987 & 58009.127 & 7724 & 0.600 & 2.73 $\pm$ 0.01 & 1568\\
    3 & 1050360105 & 58009.164 & 58009.560 & 7182 & 0.607 & 2.41 $\pm$ 0.01 & 5344\\
    4 & 1050360105 & 58009.806 & 58009.944 & 6623 & 0.631 & 1.85 $\pm$ 0.01 & 2352\\
    5 & 1050360106 & 58009.999 & 58010.524 & 6724 & 0.634 & 1.83 $\pm$ 0.01 & 6480\\
    6 & 1050360107 & 58011.864 & 58011.940 & 7843 & 0.626 & 2.15 $\pm$ 0.01 & 1760\\
    7 & 1050360109 & 58013.279 & 58013.731 & 10508 & 0.624 & 3.34 $\pm$ 0.01 & 2064\\
    8 & 1050360111 & 58015.274 & 58015.671 & 16241 & 0.641 & 9.03 $\pm$ 0.03 & 2096\\
    9 & 1050360112 & 58016.238 & 58016.957 & 16678 & 0.642 & 7.52 $\pm$ 0.05 & 2752\\
    10 & 1050360113 & 58017.010 & 58017.084 & 16367 & 0.630 & 7.59 $\pm$ 0.05 & 1664\\
    11 & 1050360113 & 58017.138 & 58017.213 & 16729 & 0.630 & 9.00 $\pm$ 0.02 & 1760\\
    12 & 1050360113 & 58017.267 & 58017.856 & 16620 & 0.630 & 6.66 $\pm$ 0.06 & 2736\\
    13 & 1130360103 & 58026.722 & 58026.816 & 13662 & 0.607 & 7.03 $\pm$ 0.03 & 3616\\
    14 & 1130360104 & 58027.753 & 58027.781 & 11816 & 0.593 & 5.43 $\pm$ 0.01 & 1808\\
    15 & 1130360105 & 58028.718 & 58028.874 & 11567 & 0.592 & 5.76 $\pm$ 0.01 & 5808\\
    16 & 1130360106 & 58029.747 & 58029.839 & 10155 & 0.590 & 6.77 $\pm$ 0.02 & 4096\\
    17 & 1130360107 & 58030.712 & 58030.800 & 9934 & 0.582 & 4.57 $\pm$ 0.01 & 2912\\
    18 & 1130360108 & 58031.355 & 58031.373 & 10064 & 0.582 & 4.85 $\pm$ 0.01 & 1584\\
    19 & 1130360108 & 58031.862 & 58031.888 & 10978 & 0.581 & 5.94 $\pm$ 0.01 & 2240\\
    20 & 1130360113 & 58036.318 & 58036.694 & 9206 & 0.556 & 5.41 $\pm$ 0.02 & 2096\\
    21 & 1130360114 & 58037.290 & 58037.938 & 7656 & 0.552 & 4.35 $\pm$ 0.01 & 4544\\
    \hline
    \end{tabular}
    \label{detail}
\end{table*}

The observations of MAXI J1535-571 that we used span from September to October 2017 from \textit{NICER}, and the ObsIDs are 1050360104 through 1050360120 and 1130360101 through 1130360114. We applied the \texttt{NICERL2} task to process each observation with the standard calibration process and screening. The NICER X-Ray Timing Instrument (XTI) delivers photons onto 56 focal plane modules (FPMs). We removed those non-operational FPMs (11, 20, 22, and 60) and data collected by FPMs 13 and 34. After that, we used \texttt{NICERL3-LC} to produce light curves with a time resolution of $0.0078125$ s and limited our events within the 1-10 keV energy range to avoid the energy redistribution below 1 keV and strong background higher than 10 keV.

\subsection{Timing analysis}
\label{time}
With a time resolution of $0.0078125$ s, the corresponding Nyquist frequency is 64 Hz. We extracted the fractional rms amplitude normalized PDS with powspec 1.0, and each interval consists of 2048 points (16s). Then, those extracted PDS were fitted with a multi-lorentized model using \texttt{Xspec} version 12.13.0c. The fitting model generally has four or five Lorentzians representing the broad-band noise component, the QPO, and the second harmonics. Each Lorentzian contains three parameters: The centroid frequency $\nu_0$, the full-width at half maximum, and the normalization.

We found that the count rate and the QPO frequency may change significantly within a single observation. To avoid averaging features of different states, for some of the observations, we divided them into several segments to maintain a relatively constant count rate and QPO frequency within each segment. We visually selected and mainly focused on those segments with clear type-C QPOs. Then, we analyzed those segments with bicoherence independently, and only kept those segments with enough exposure time (see Section \ref{stat}). The detailed information of the selected segments is listed in Table \ref{detail}.

\subsection{Bicoherence statistical method}
\label{stat}

Our work primarily focuses on the computation of bispectrum and bicoherence. The bispectrum is a method to analyze the coupling of phases among Fourier components and inspect the non-linearity of the light curves. A time series is divided into \textit{K} intervals, and the bispectrum of this time series is:
\begin{equation}
    B(k, l)=\frac{1}{K}\sum_{i=0}^{K-1}X_i(k)X_i(l)X_i^*(k+l)
    \label{bispectrum},
\end{equation}
where $X_i(f)$ is the Fourier transform of the \textit{i}th intervals of the time-series at frequency $f$, and $X_i^*(f)$ is the conjugate of $X_i(f)$. Since bispectrum is a complex number, it can be denoted by a magnitude with a phase in the complex plane, and this phase is called biphase.

The bicoherence, ranging from 0 to 1, is the normalized magnitude of the bispectrum, given by:
\begin{equation}
    b^2=\frac{|\sum X_i(k)X_i(l)X_i^*(k+l)|^2}{\sum|X_i(k)X_i(l)|^2\sum|X_i(k+l)|^2}
    \label{bicoherence}.
\end{equation}
This normalization is proposed by \cite{4317207}. The value of $b^2$ indicates the phase coupling of the three frequencies, k, l and k+l, that is, the consistency of biphase. If the coupling is strong, the biphase stays approximately constant over the observation, and the value of $b^2$ approaches 1. If the coupling is weak, the phase distributes randomly, and the value of $b^2$ tends to be 0 when K is large enough.

After computing the bicoherence values of different observations at different $f_1$ and $f_2$ (that is, the frequency k and l in equation \ref{bispectrum} and \ref{bicoherence}), we used $f_1$ and $f_2$ as axes and colors to denote different values of log$b^2$ at different points. We found different phenomenological patterns in bicoherence. To classify those bicoherence patterns, we use the classification, which differentiates them as 'web', 'cross', and 'hypotenuse', as mentioned in \cite{Maccarone_2011} and \cite{Arur_2019}.

To improve the statistical reliability of our results, the segments need to be long enough to get sufficient cycles for our analysis \citep{Arur2020MNRAS.491..313A}. For this reason, we exclude those segments with exposure time less than 1.5 ks (corresponding to about 94 intervals). We also chose 16 s as the length of each interval so that the frequency resolution is 0.0625 Hz, which is enough for observations with a QPO frequency of $\gtrsim$ 2 Hz, and we can get as many intervals as possible.

\section{Result}
\label{result}
\subsection{Light curve and HID}
\label{lchid}
\begin{figure}
    \begin{minipage}{0.48\textwidth}
		\includegraphics[width=1\textwidth]{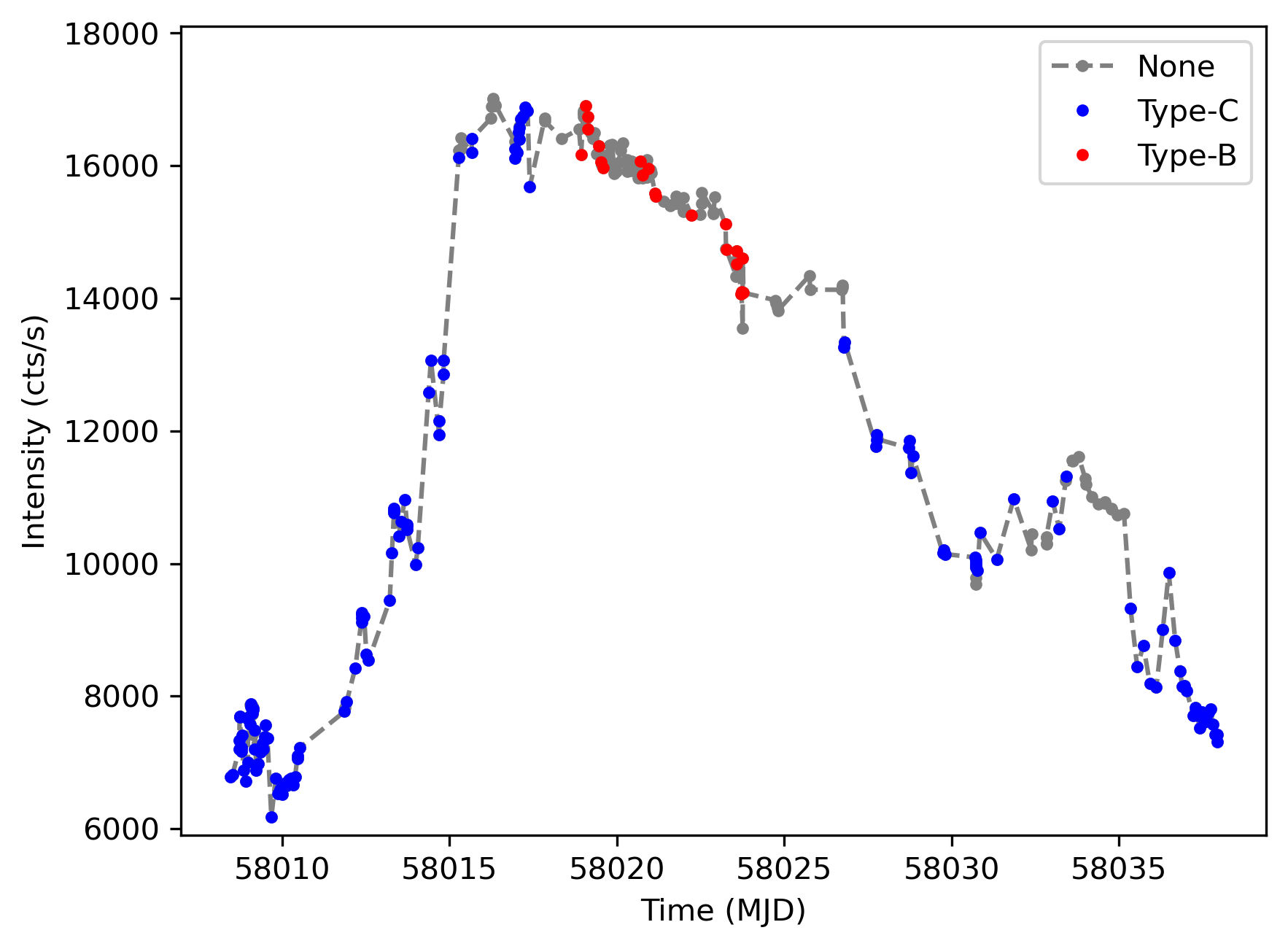}
	\end{minipage}
	\begin{minipage}{0.48\textwidth}
		\includegraphics[width=1\textwidth]{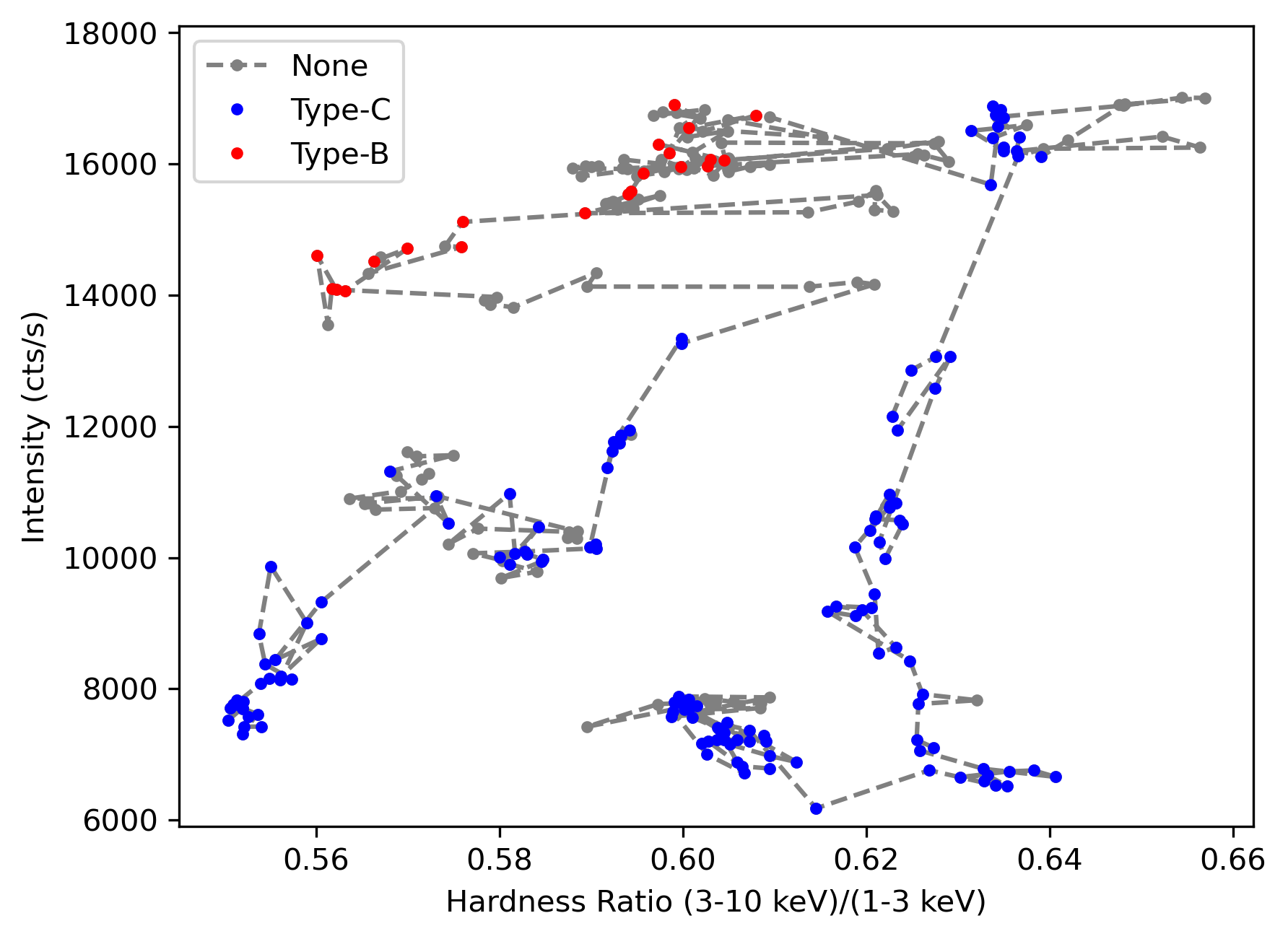}
	\end{minipage}
	\caption{Light curve (top panel) and hardness-intensity diagram (bottom panel) of MAXI J1535-571 during the outburst in 2017. Every data point represents a GTI. The grey, blue, and red points correspond to the GTIs without QPO, with type-C QPO, and with type-B QPO, respectively. The data points are connected with a dash grey line in chronological order.}
	\label{hid}
\end{figure}

The light curve and HID diagram of MAXI J1535-571 during the outburst in 2017 are shown in Fig. \ref{hid}. Every data point represents a GTI (longer than 16 s). The source count rate in the 1-10 keV band is used as our intensity, and our hardness ratio is the ratio of the source count rate in the 3-10 keV band and the 1-3 keV band. We marked the GTIs where type-C QPO occurs as blue points and those with type-B QPO as red points. GTIs without QPO are grey points. We also used a dashed grey line to connect the data points chronologically.

The source moves in an anticlockwise direction in the HID diagram. At the beginning of the outburst, the source stays in LHS with a count rate of $\sim$8000 cts/s and a hardness ratio of $\sim$0.6. Then, the source count rate rises rapidly to $\sim$16000 cts/s, while the hardness ratio remains $\sim$ 0.62. With the disappearance of type-C QPO and the occurrence of type-B QPO from time to time, the source moves from LHS to a softer state. The count rate falls to $\sim$14000 cts/s, and the hardness ratio decreases to $\sim$ 0.56. After the type-C QPO reappears, the count rate continuously goes down, and the hardness ratio drops from $\sim$0.6 to $\sim$0.56.

\subsection{PDS}
\label{pds}

\begin{figure}
    \begin{minipage}{0.48\textwidth}
		\includegraphics[width=1\textwidth]{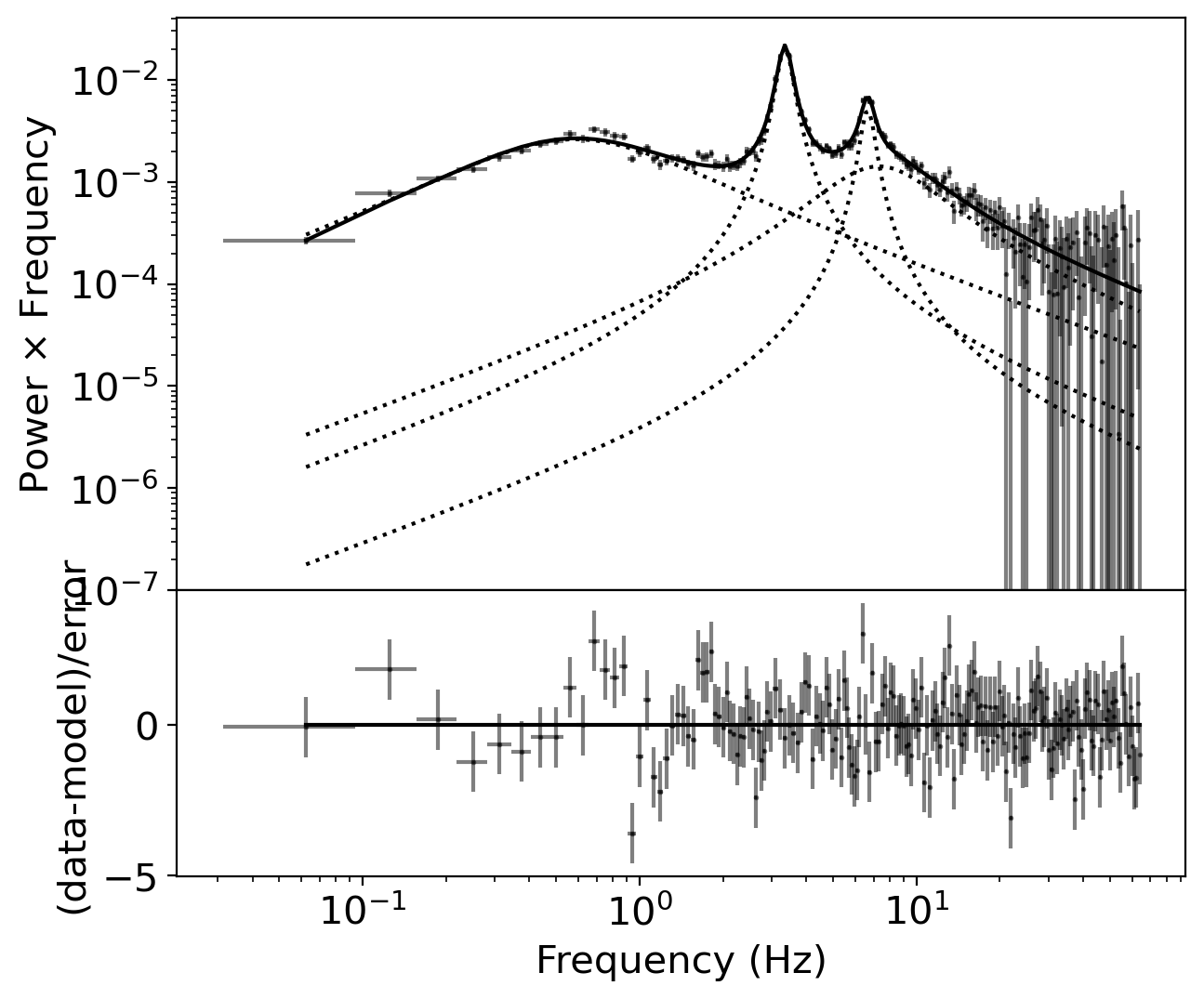}
	\end{minipage}
	\begin{minipage}{0.48\textwidth}
		\includegraphics[width=1\textwidth]{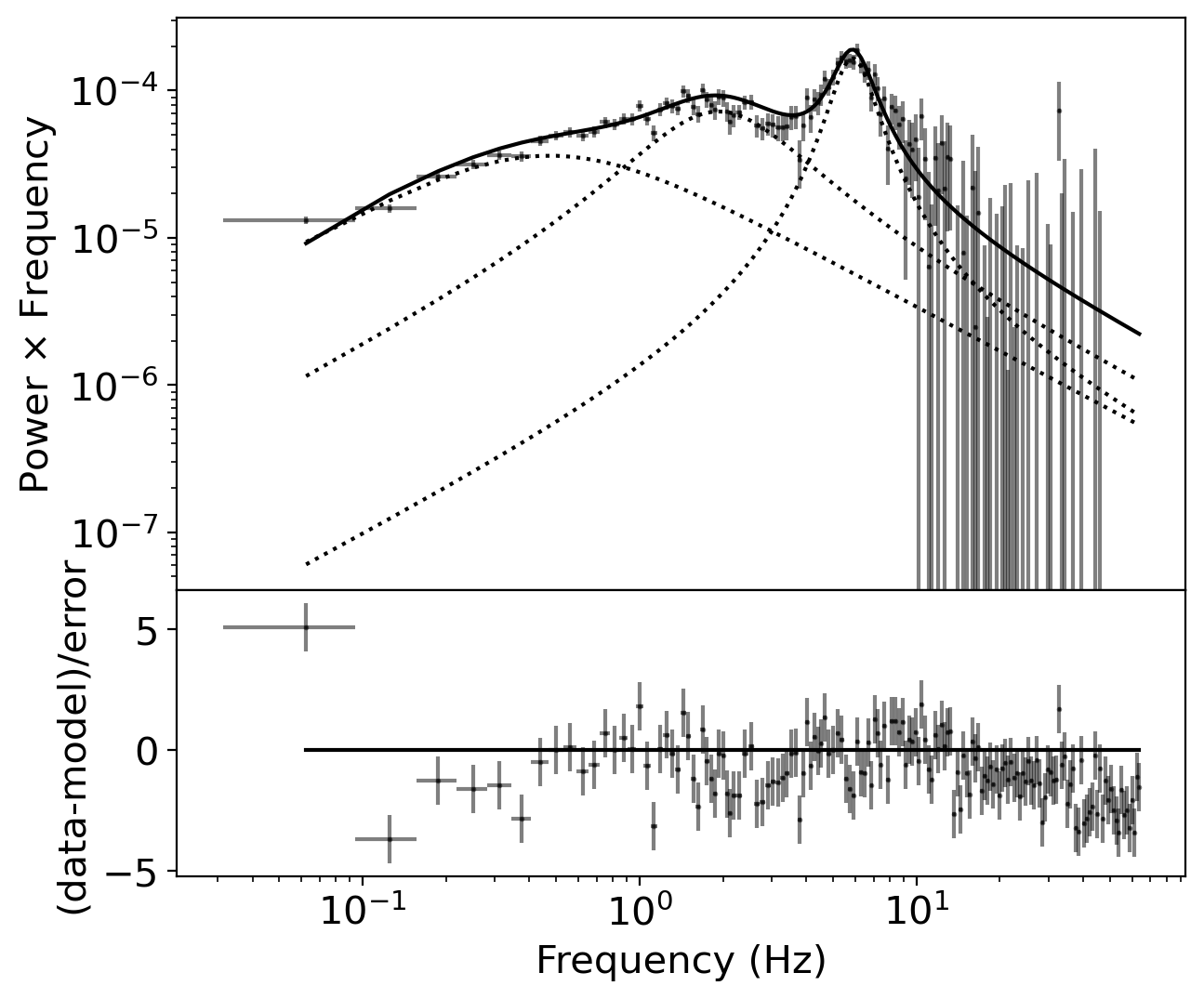}
	\end{minipage}
	\caption{{\bf Top panel: } the PDS of Obs no.7 in Table \ref{detail} as an example of type-C QPO. {\bf Bottom panel: } the PDS produced from the light curve combines all the gtis with type-B QPOs. The PDS is fitted with three or four Lorentzians, followed by a residual plot below.}
	\label{qpo}
\end{figure}

In Fig. \ref{qpo}, we show representative PDS of observations with type-C QPO and type-B QPO accompanied by their best fits. We took the PDS of Obs no.7 in Table \ref{detail} as an example of type-C QPO with residuals plotted below. We fitted this PDS with four Lorentzians. The type-C QPO has a narrow peak with a high rms amplitude and always comes with a second harmonic. 
It is worth pointing out that the type-B QPO example is not from a single observation. Since type-B QPO is weak and unstable in this system, we combined all the GTIs with type-B QPO into a single light curve to generate the PDS, revealing a relatively strong type-B QPO. This PDS is fitted with three Lorentzians. Compared to type-C QPO, type-B QPO's peak is relatively broad, and the rms amplitude is low. We do not find a second harmonic in this PDS.

\subsection{bicoherence patterns}
\label{bico_patt}

During the softening process of MAXI J1535-571, a transition in bicoherence is observed. At the beginning of the outburst, when type-C QPO first exists, a 'web' pattern is found. Then the type-C QPO disappears, and the type-B QPO shows up. After the period of type-B QPO, type-C QPO appears again, and the pattern turns into a 'hypotenuse'.

\begin{figure*}
\centering
    \subfigure[Obs no.1]{
    \begin{minipage}[]{0.45\linewidth}
     \includegraphics[width=\textwidth,height=0.7\textwidth]{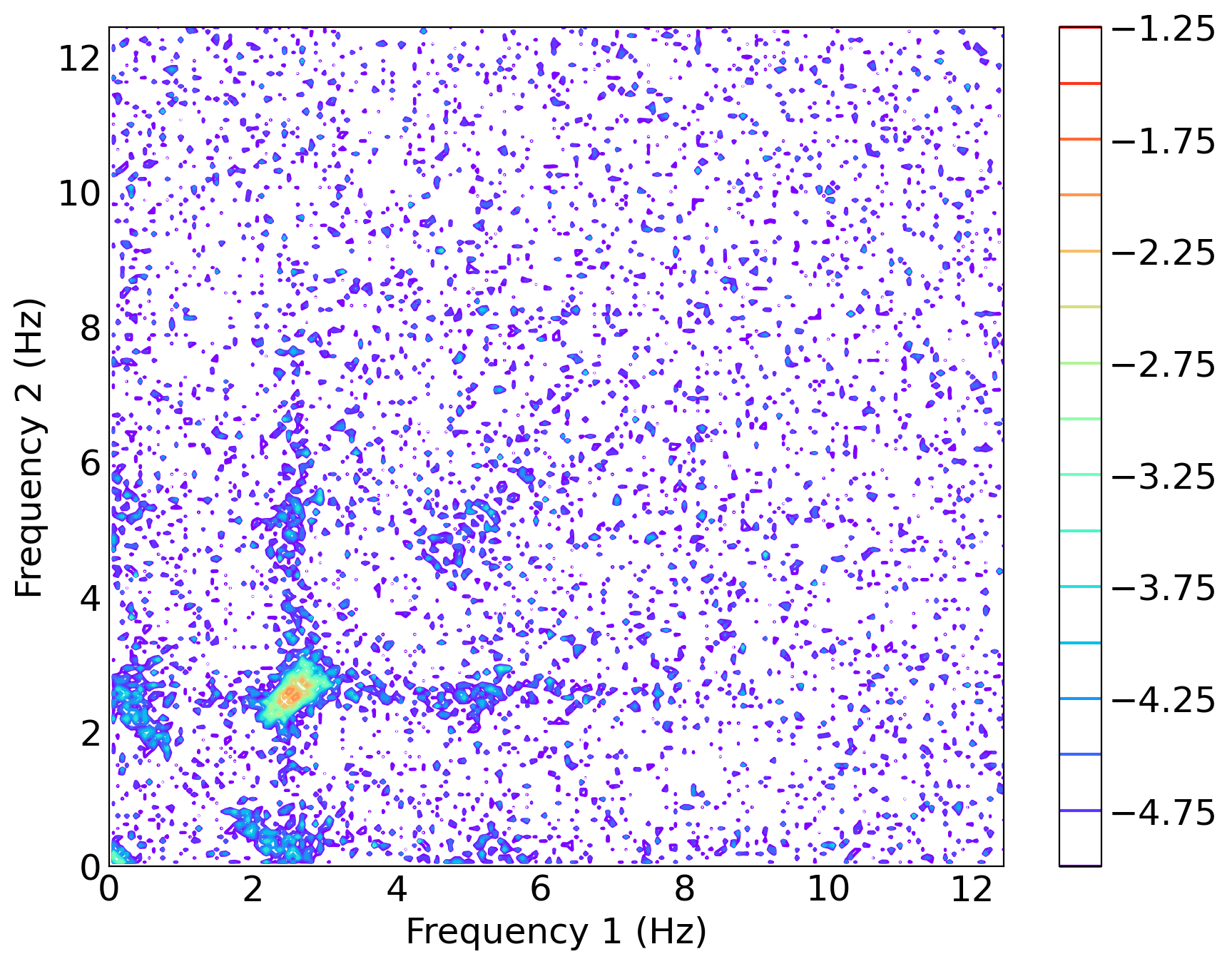}
     \flushleft
     \includegraphics[width=0.9\textwidth,height=0.6\textwidth]{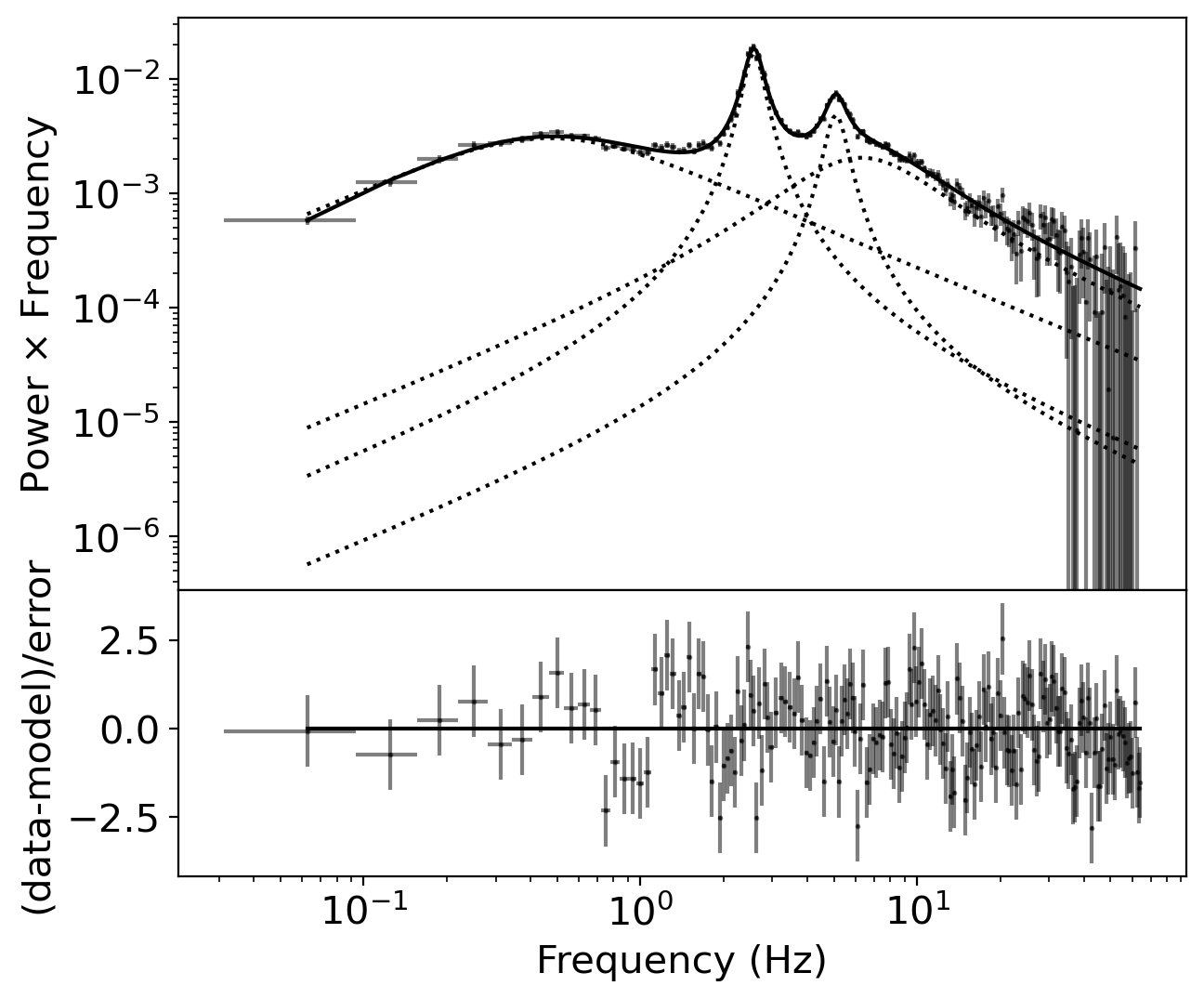}
     \end{minipage}
    }\hfill
    \subfigure[Obs no.5]{
    \begin{minipage}[]{0.45\linewidth}
     \includegraphics[width=\textwidth,height=0.7\textwidth]{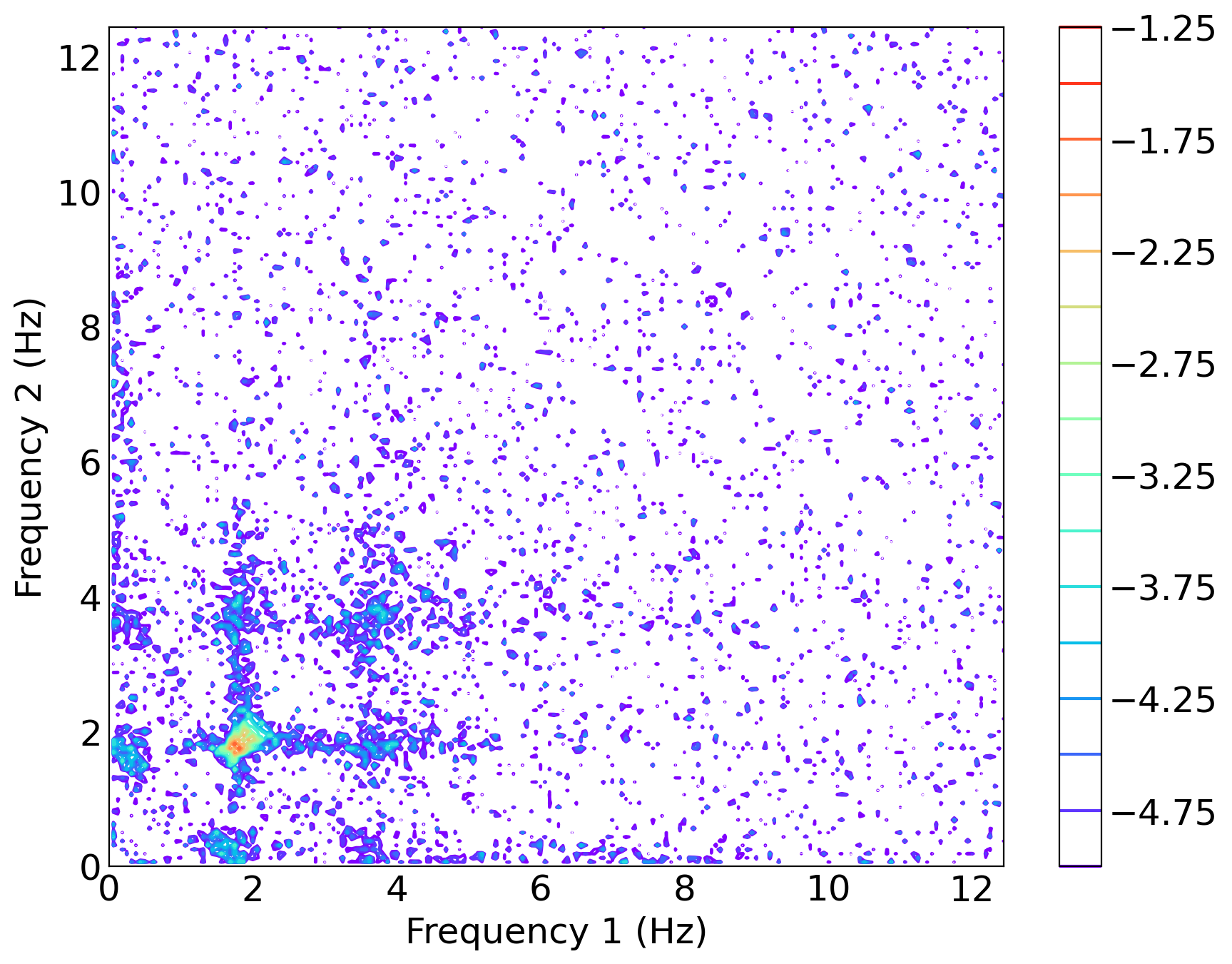}
     \flushleft
     \includegraphics[width=0.9\textwidth,height=0.6\textwidth]{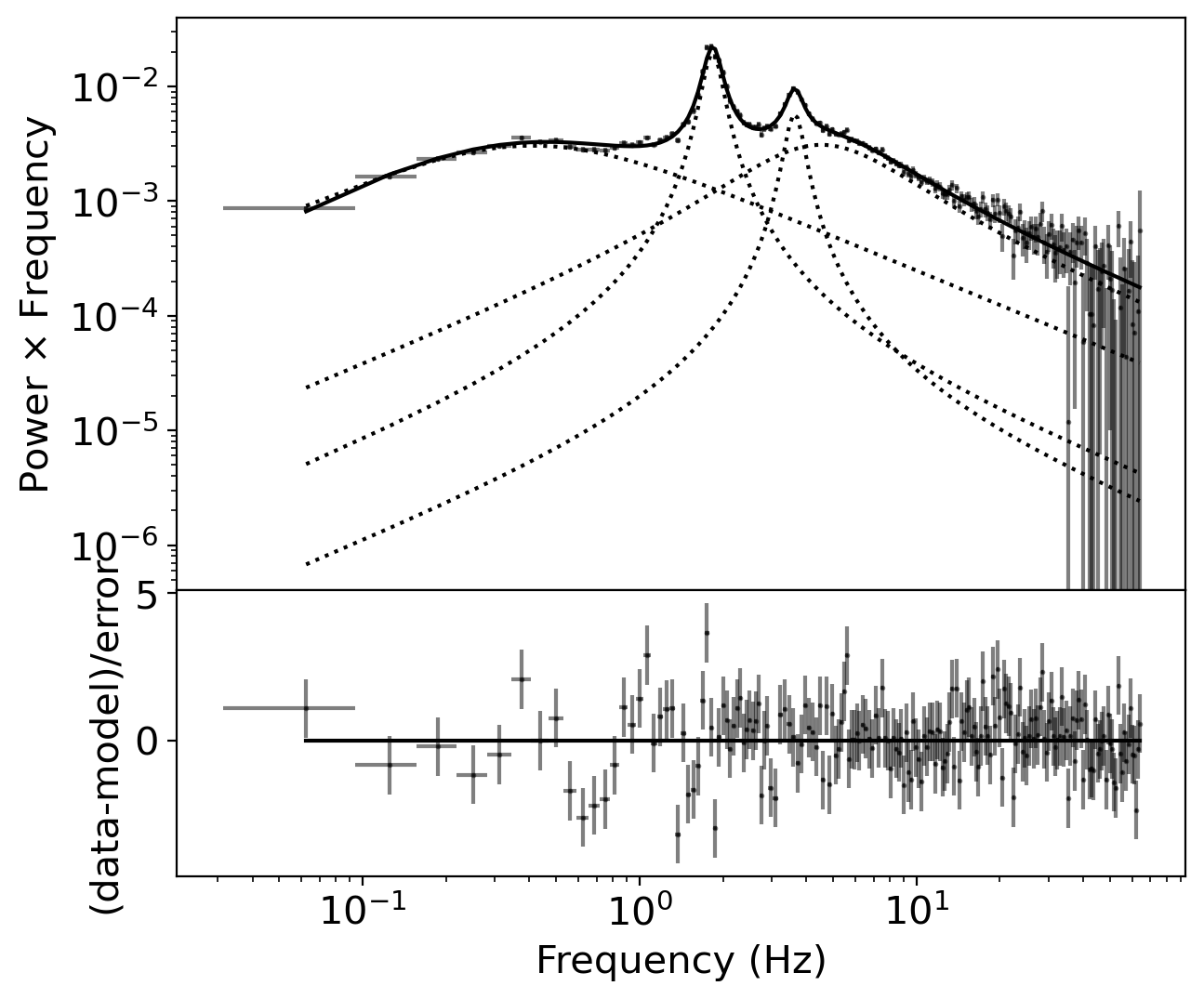}
      \end{minipage}
    }\hfill
    \subfigure[Obs no.6]{
    \begin{minipage}[]{0.45\linewidth}
     \includegraphics[width=\textwidth,height=0.7\textwidth]{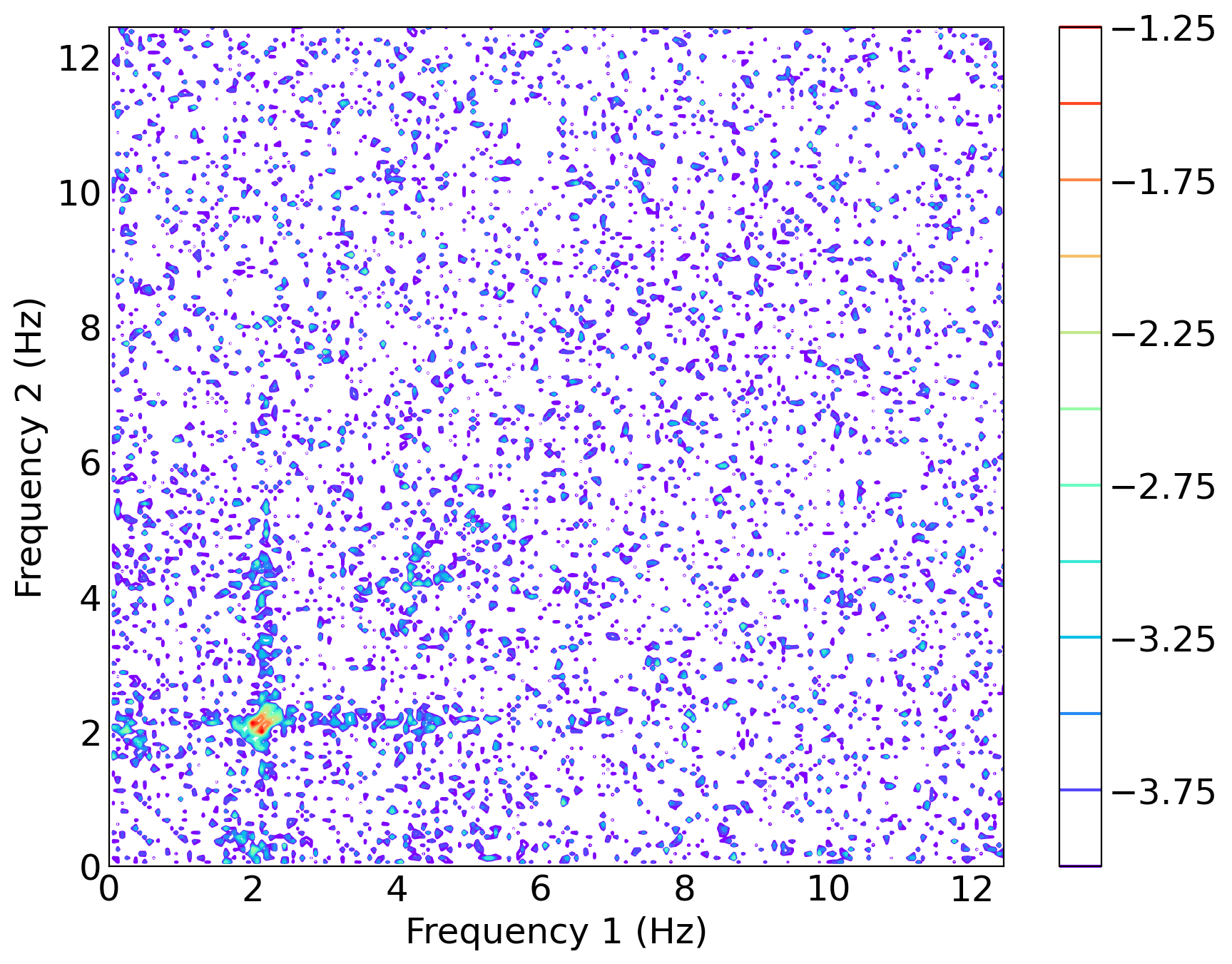}
     \flushleft
     \includegraphics[width=0.9\textwidth,height=0.6\textwidth]{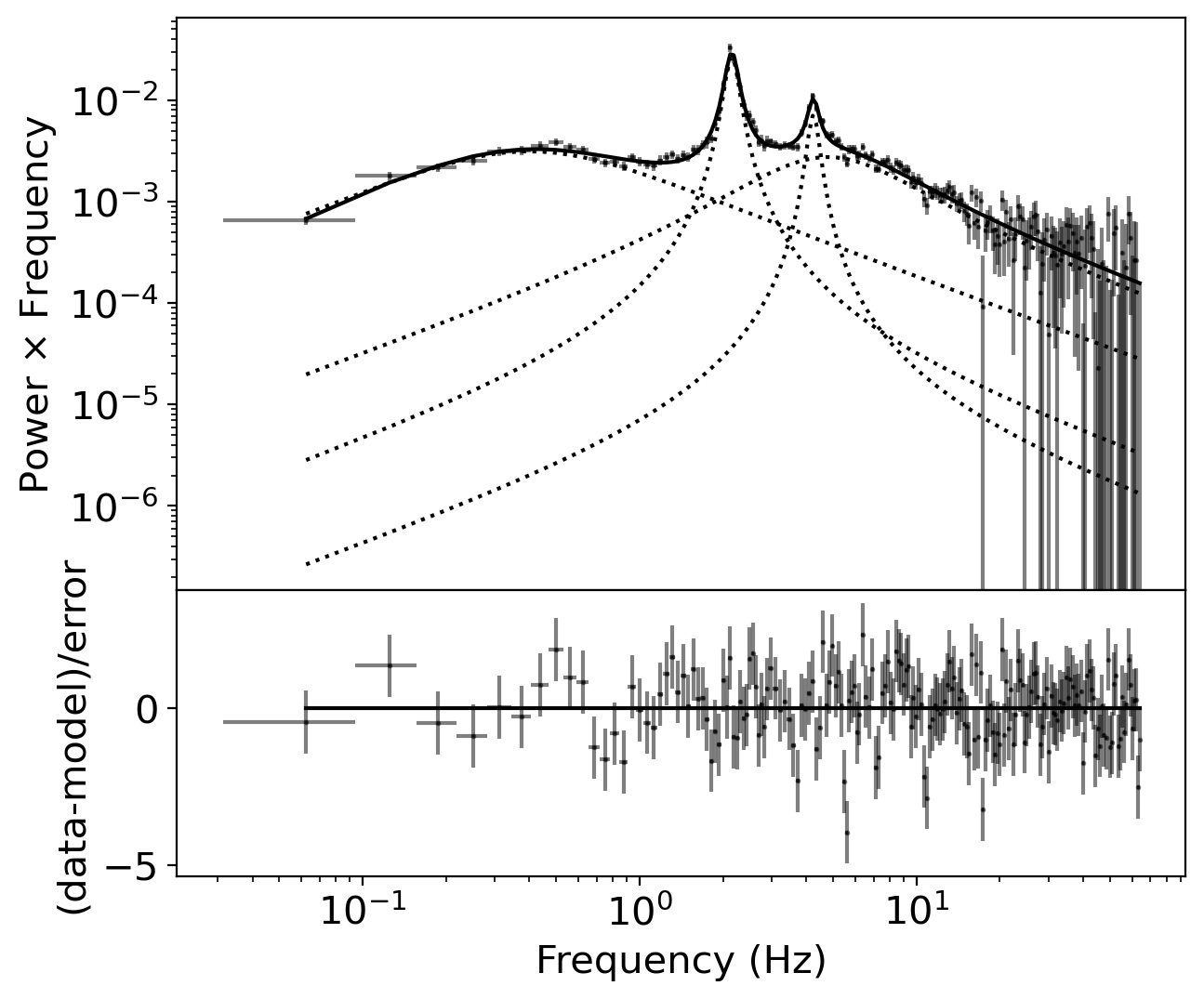}
      \end{minipage}
    }\hfill
    \subfigure[Obs no.7]{
    \begin{minipage}[]{0.45\linewidth}
     \includegraphics[width=\textwidth,height=0.7\textwidth]{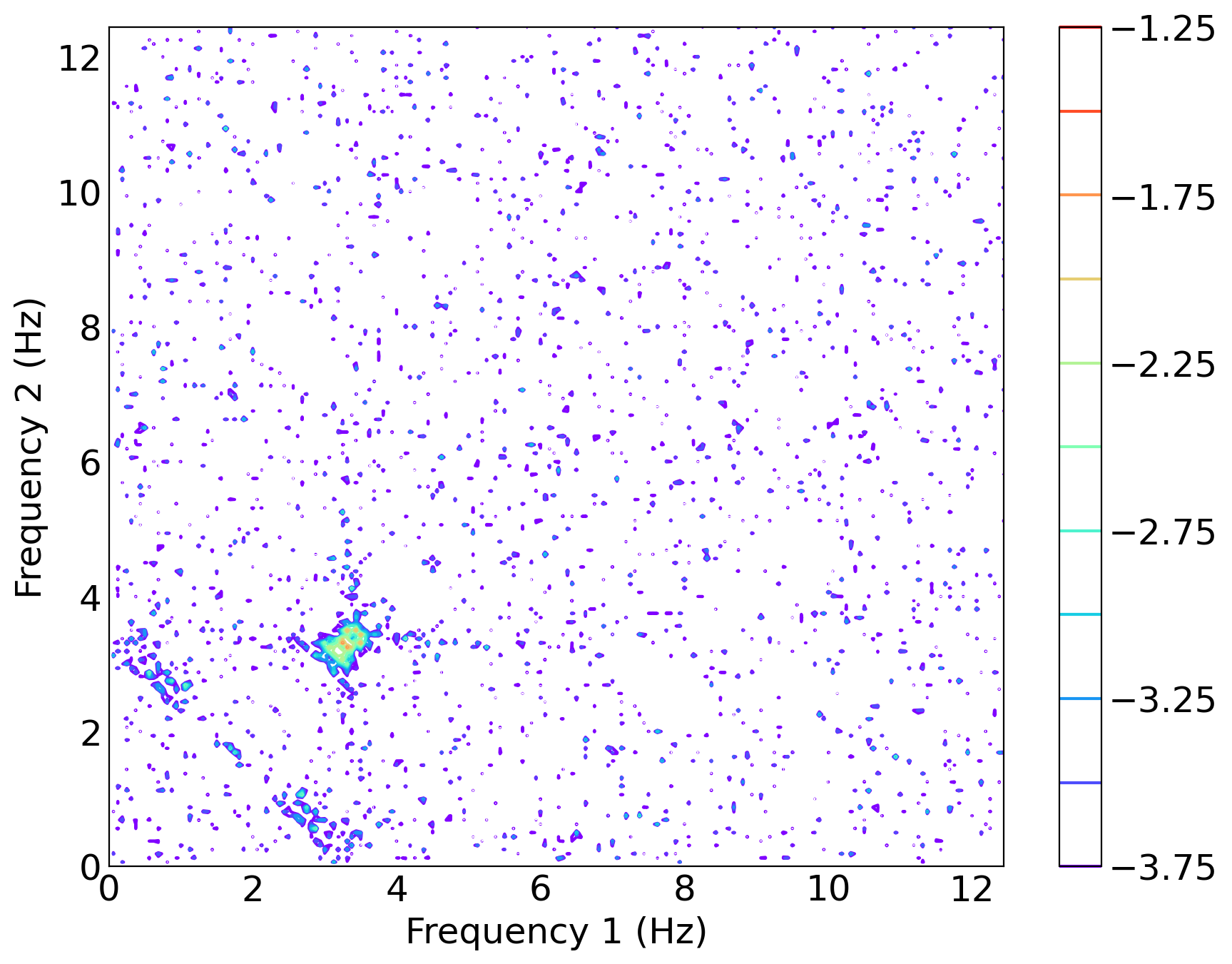}
     \flushleft
     \includegraphics[width=0.9\textwidth,height=0.6\textwidth]{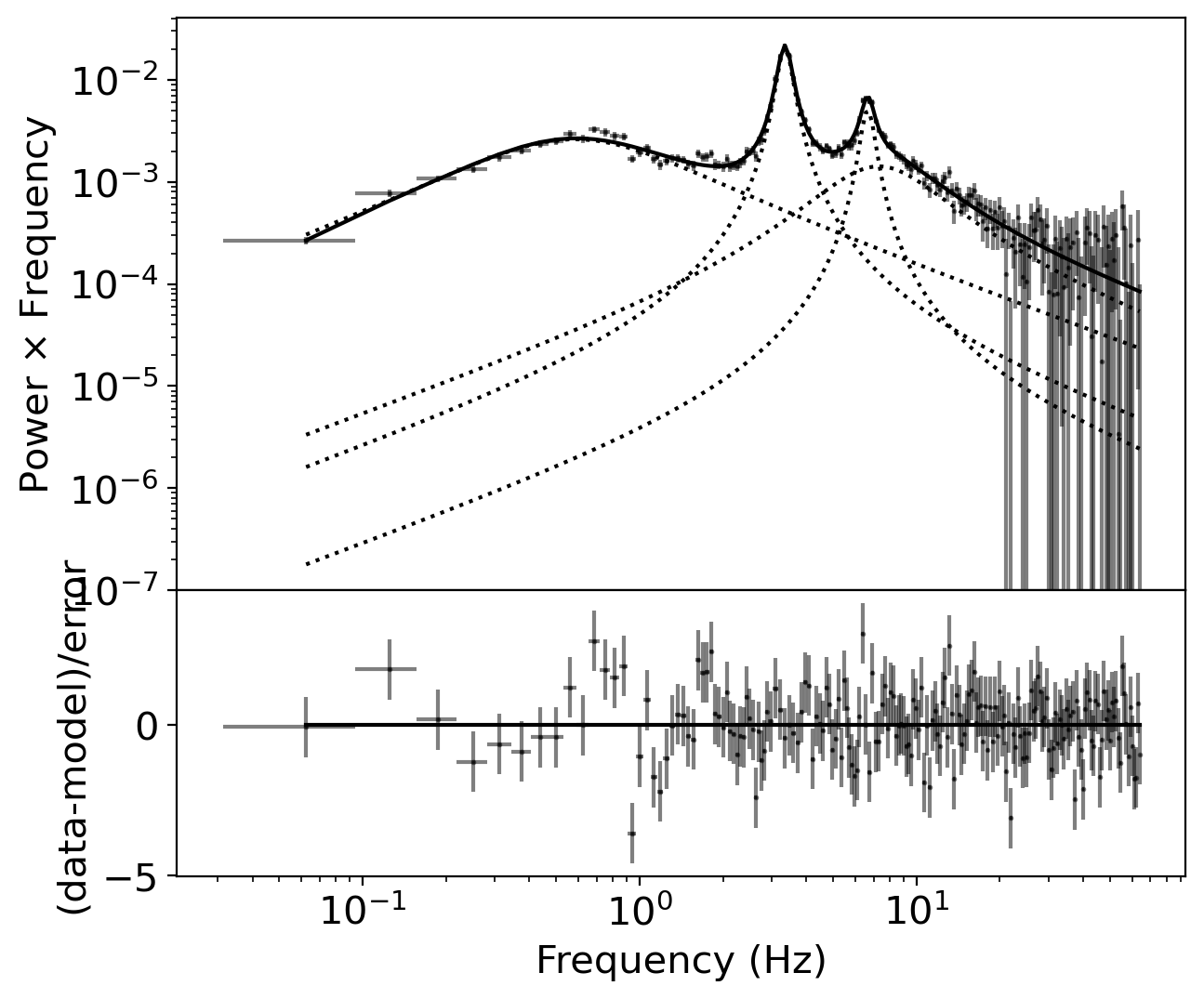}
      \end{minipage}
    }
        \caption{The bicoherence patterns and PDS when the type-C QPO first appears. During this period, the bicoherence patterns are 'web' patterns, that is, the high bicoherence is seen where one frequency equals to that of QPO (and harmonics) and two frequency components add to the frequency of QPO. The PDS is fitted with four Lorentzians, followed by a residual plot below.}
    \label{per1}
\end{figure*}

In Fig. \ref{per1}, we show examples of the bicoherence patterns when the type-C QPO first appeared. During this period, We can see high bicoherence where one frequency equals to that of the QPO (and harmonics), i.e. the horizontal and vertical lines in the patterns. High bicoherence is also found where two frequency components add to the frequency of the QPO, i.e. the diagonal streaks in the patterns. These patterns are categorized as 'web' patterns. Additionally, the bicoherence is notably stronger where both $f_1$ and $f_2$ equal $f_{QPO}$, providing evidence of a strong coupling between the QPO and its second harmonic. This phenomenon is also an indicator of the presence of higher harmonics in some segments. We also present the PDS for the corresponding segments in Fig. \ref{per1}, fitted with four Lorentzians and followed by a residual plot below. The components of PDS during this period remain consistent, comprising a QPO, a second harmonic and two broadband noises.

\begin{figure*}
\centering
    \subfigure[Obs no.14]{
    \begin{minipage}[]{0.45\linewidth}
     \includegraphics[width=\textwidth,height=0.7\textwidth]{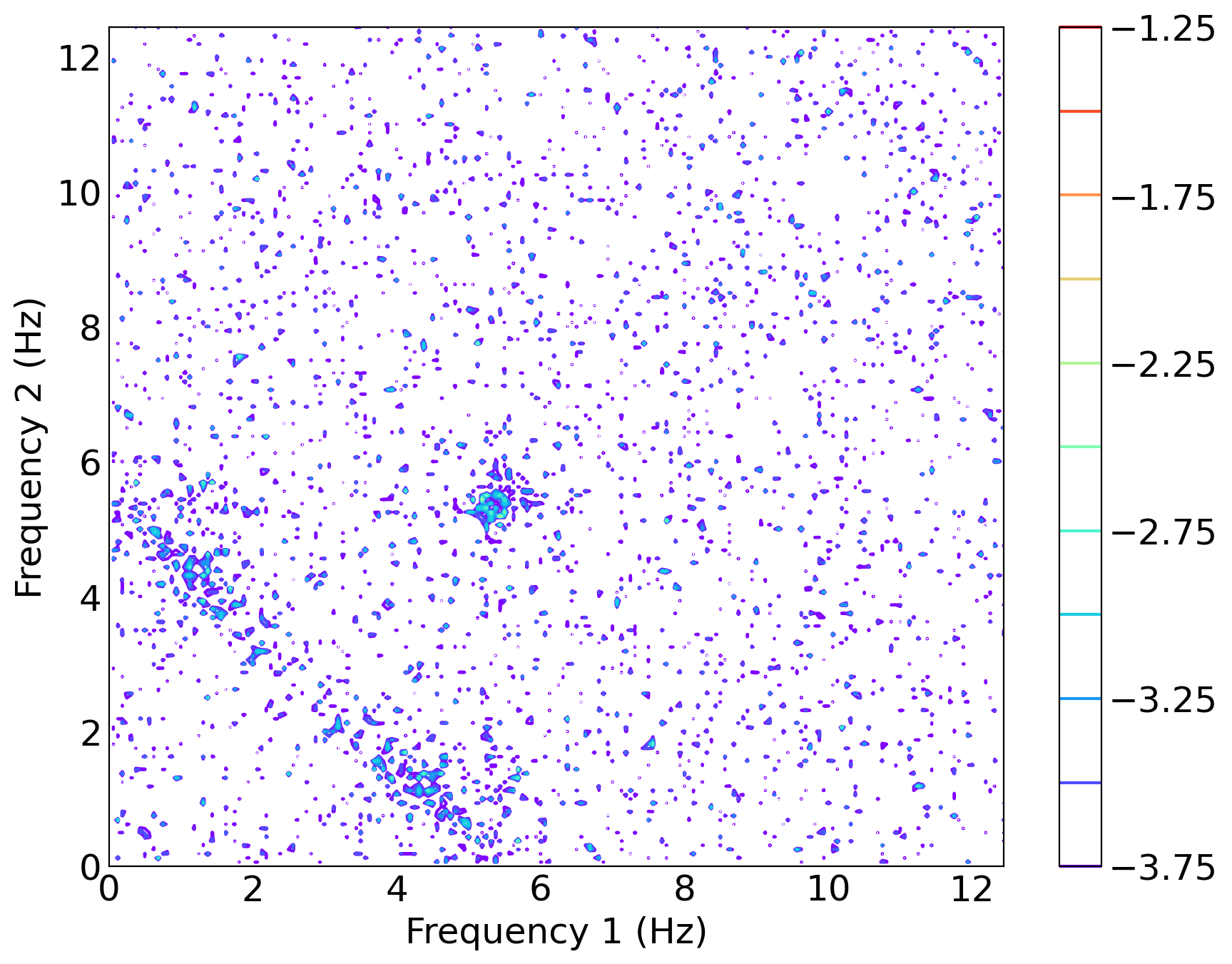}
     \flushleft
     \includegraphics[width=0.9\textwidth,height=0.6\textwidth]{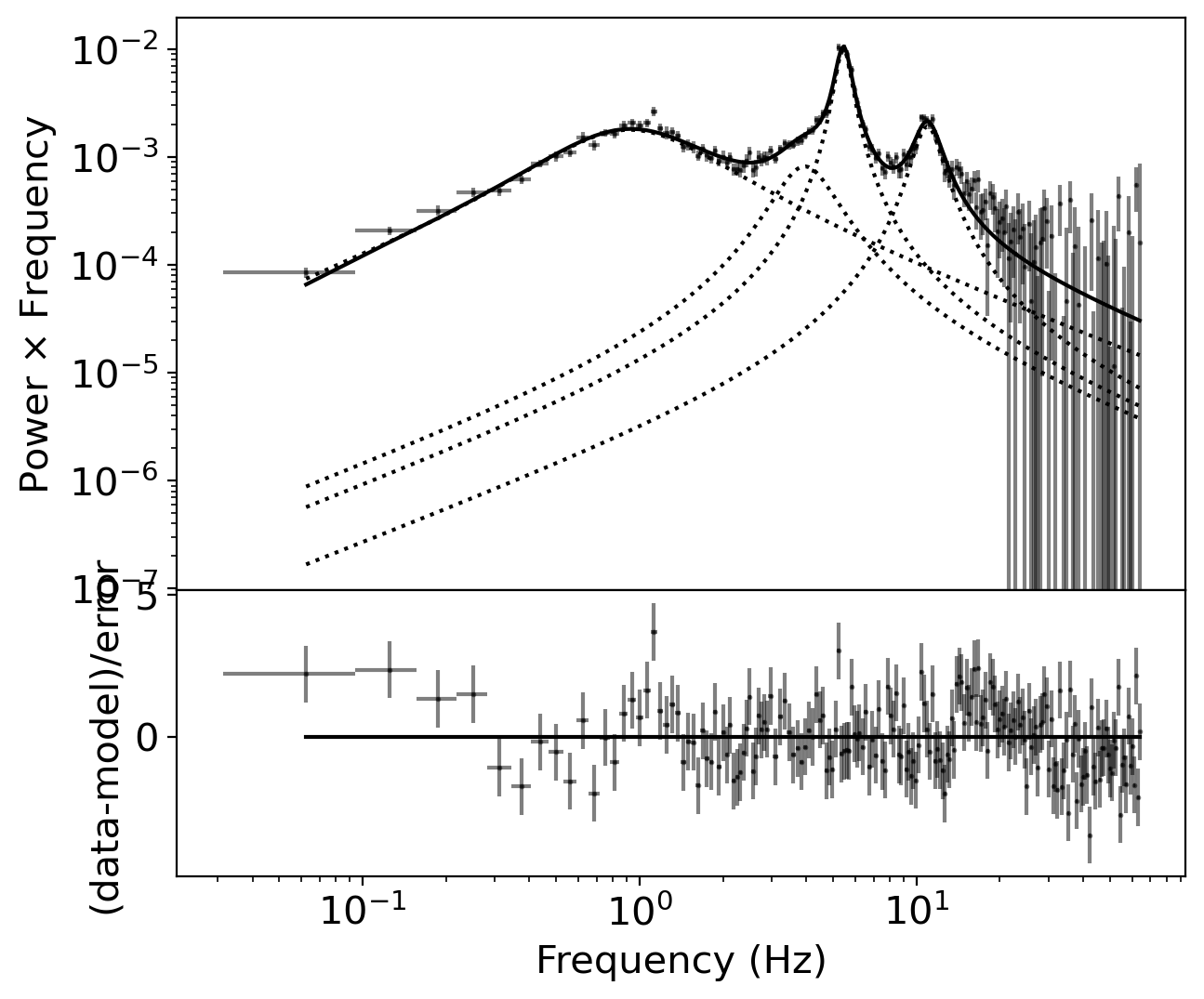}
      \end{minipage}
    }\hfill
    \subfigure[Obs no.17]{
    \begin{minipage}[]{0.45\linewidth}
     \includegraphics[width=\textwidth,height=0.7\textwidth]{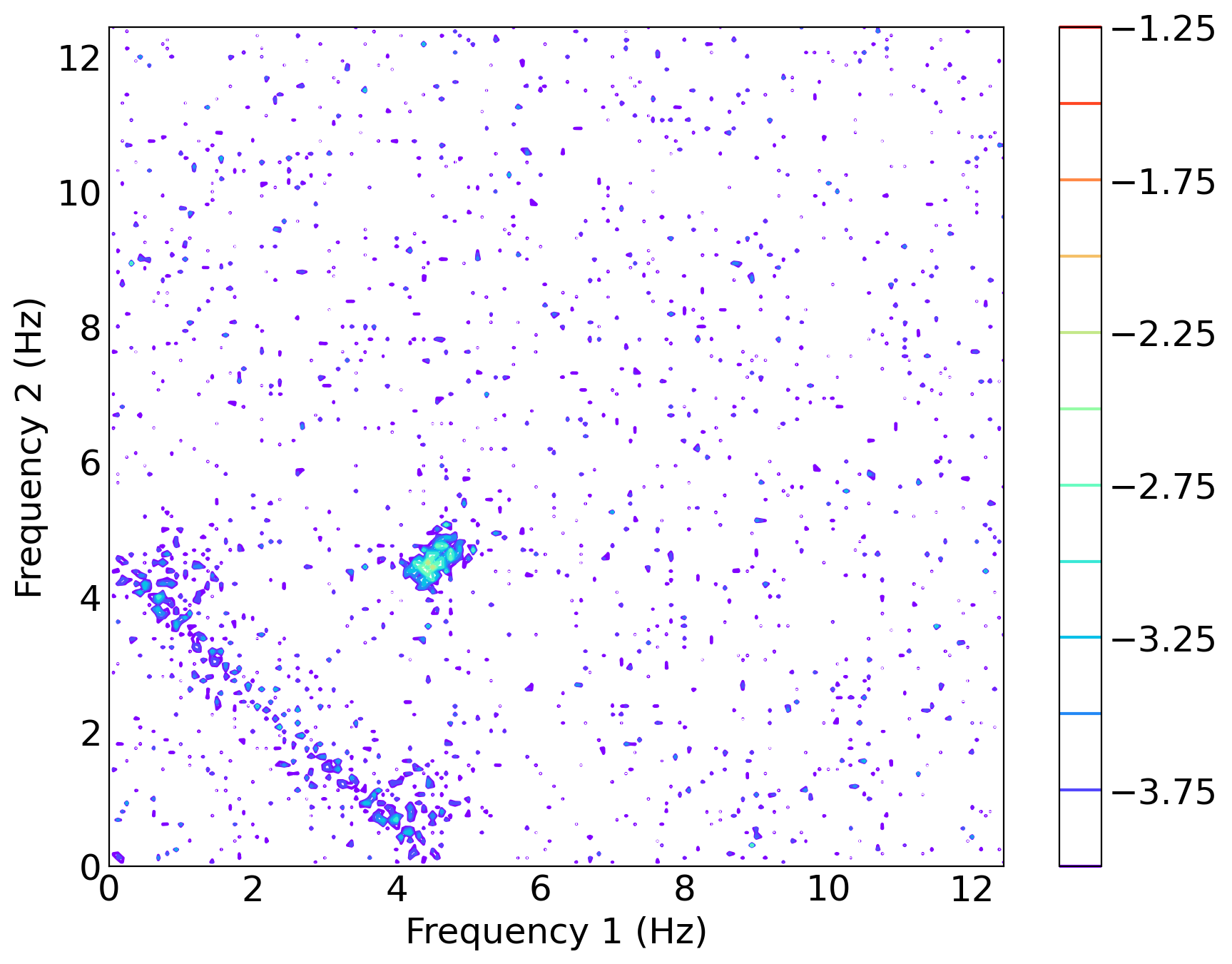}
     \flushleft
     \includegraphics[width=0.9\textwidth,height=0.6\textwidth]{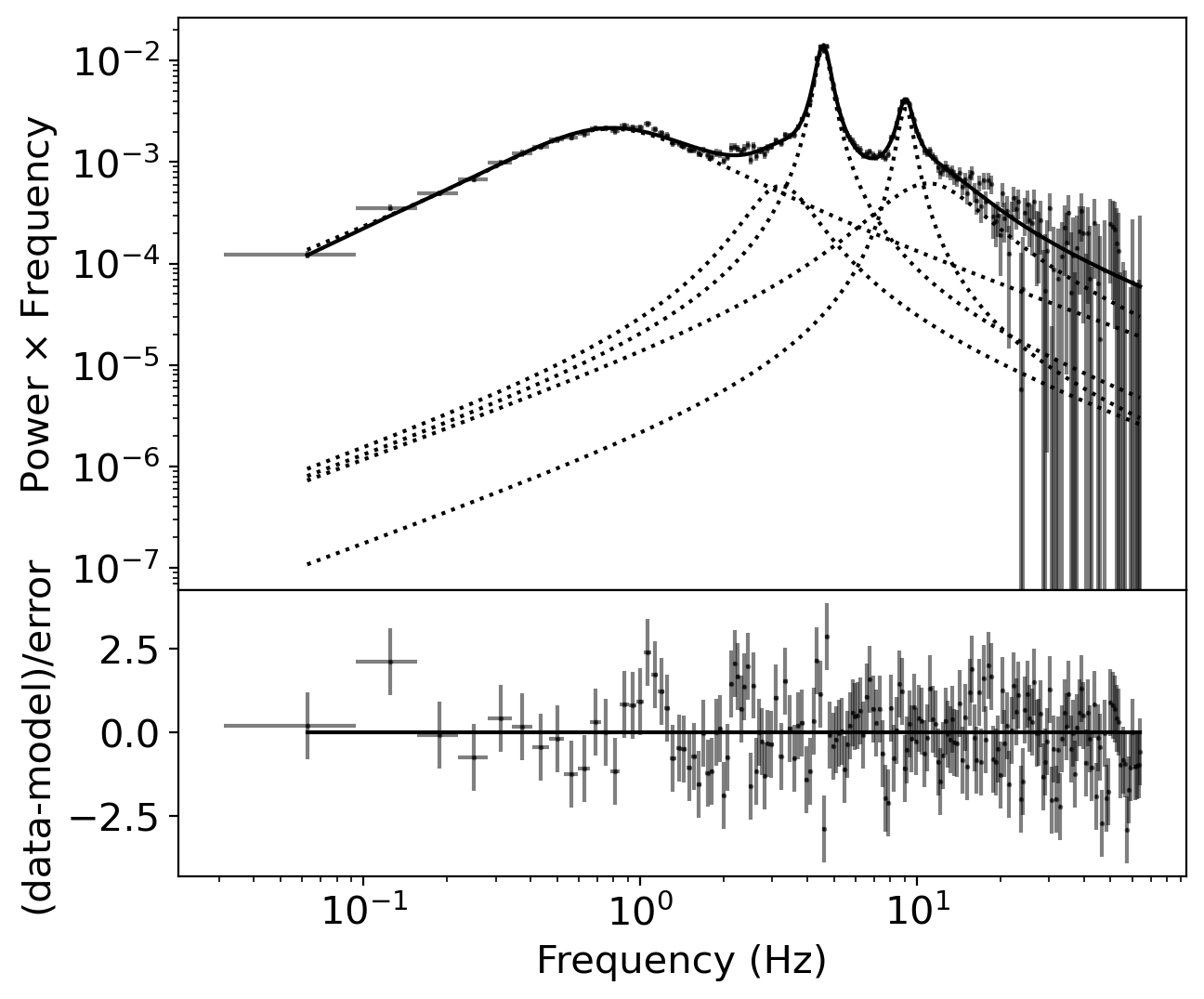}
      \end{minipage}
    }\hfill
    \subfigure[Obs no.20]{
    \begin{minipage}[]{0.45\linewidth}
     \includegraphics[width=\textwidth,height=0.7\textwidth]{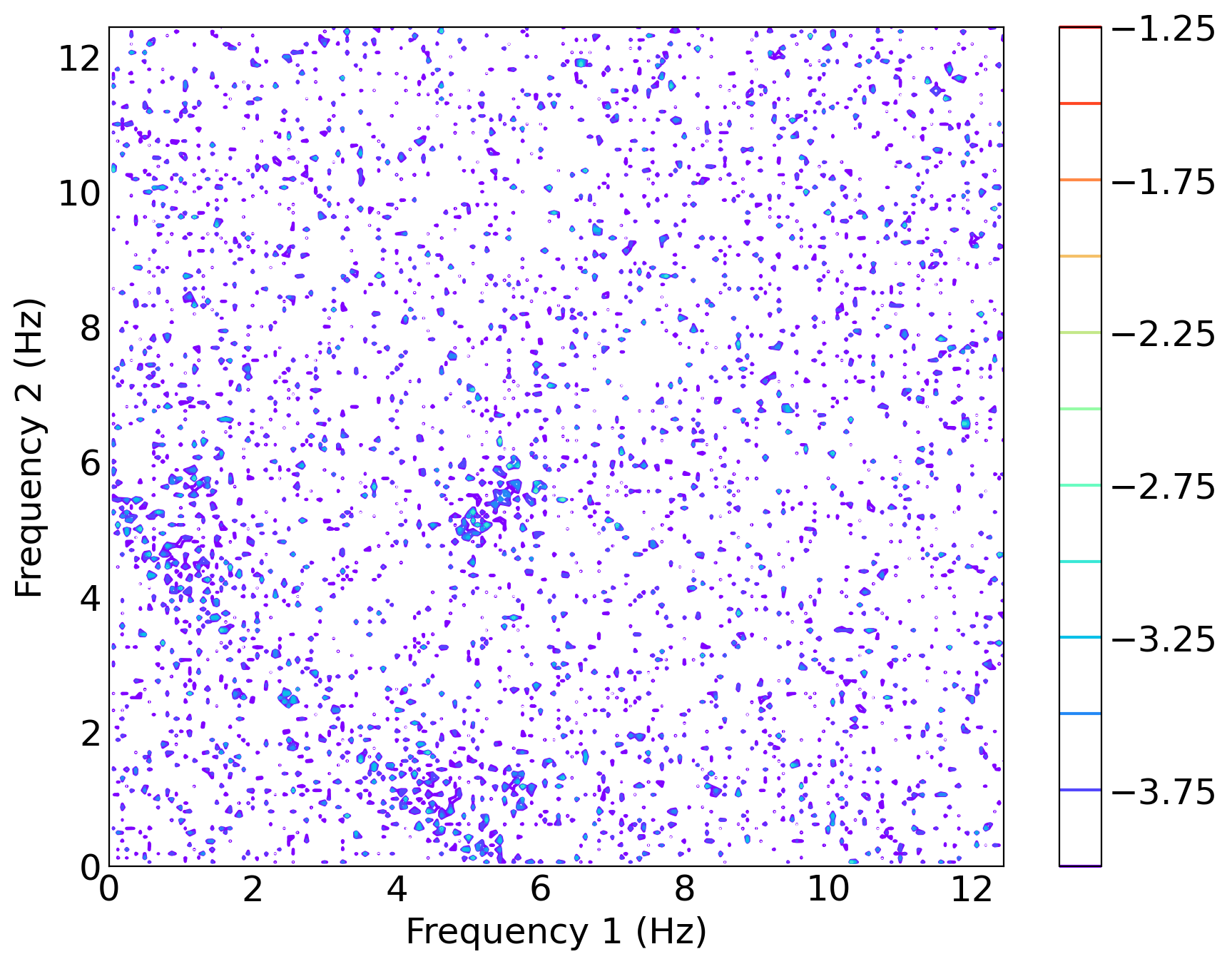}
     \flushleft
     \includegraphics[width=0.9\textwidth,height=0.6\textwidth]{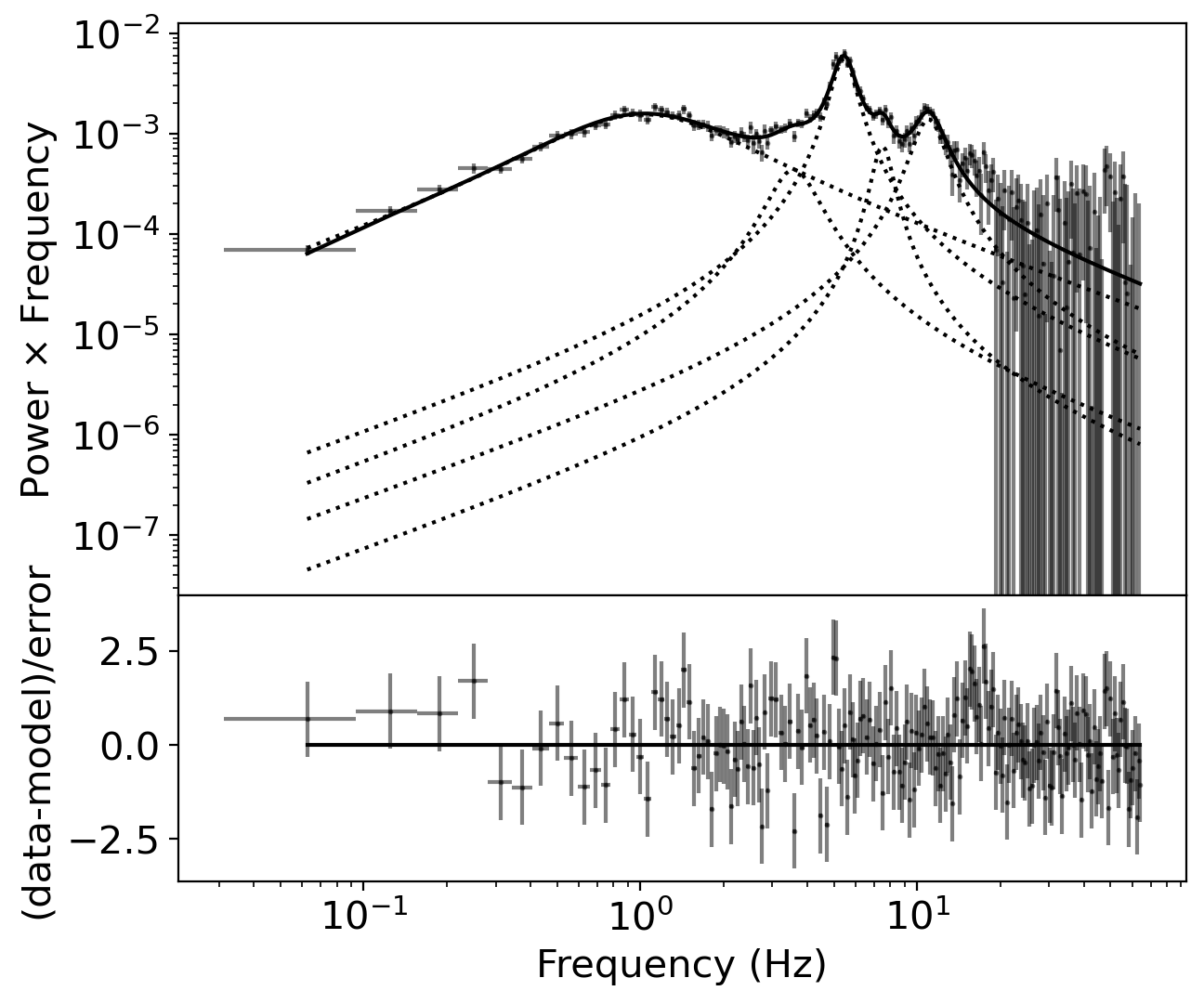}
      \end{minipage}
    }\hfill
    \subfigure[Obs no.21]{
    \begin{minipage}[]{0.45\linewidth}
     \includegraphics[width=\textwidth,height=0.7\textwidth]{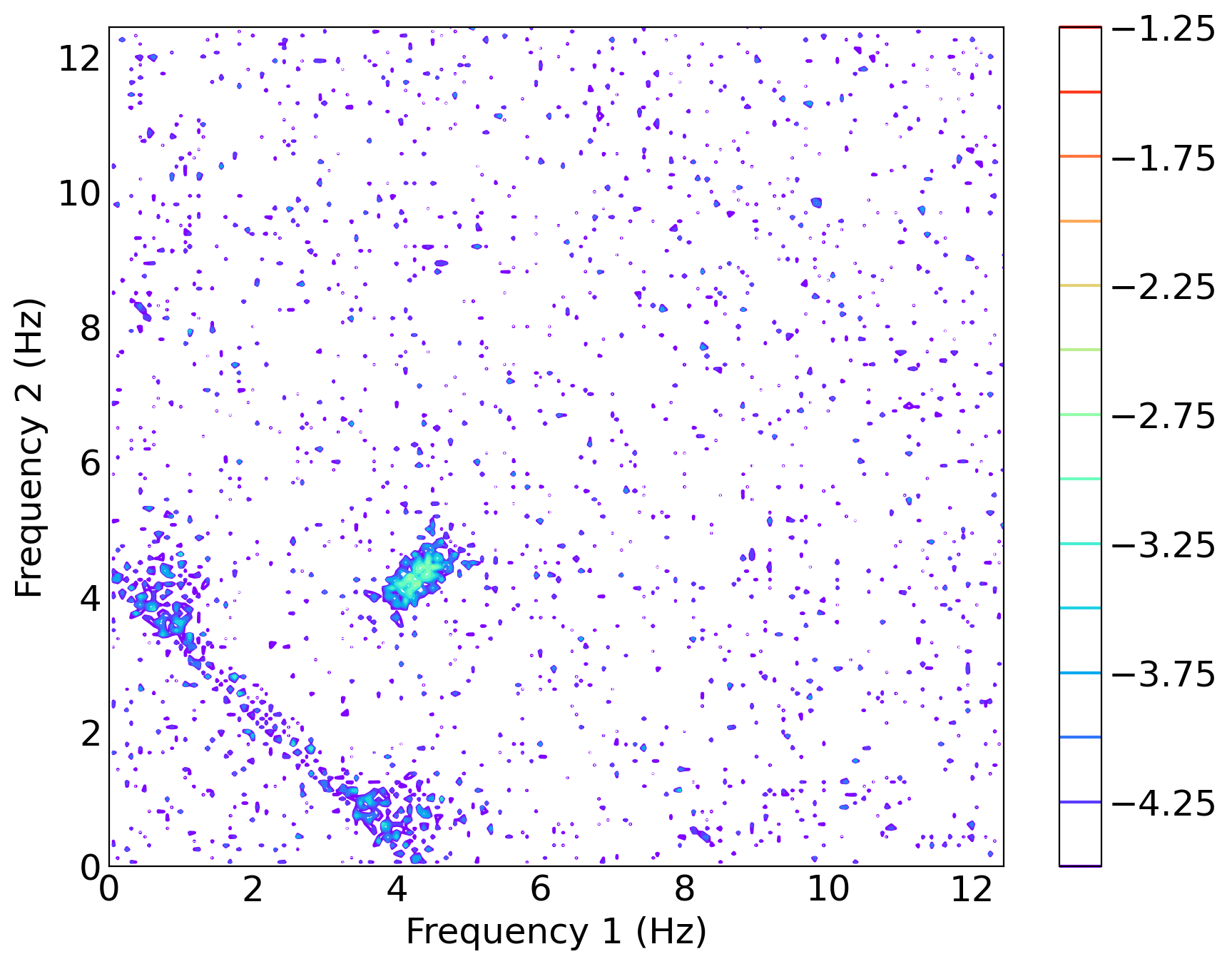}
     \flushleft
     \includegraphics[width=0.9\textwidth,height=0.6\textwidth]{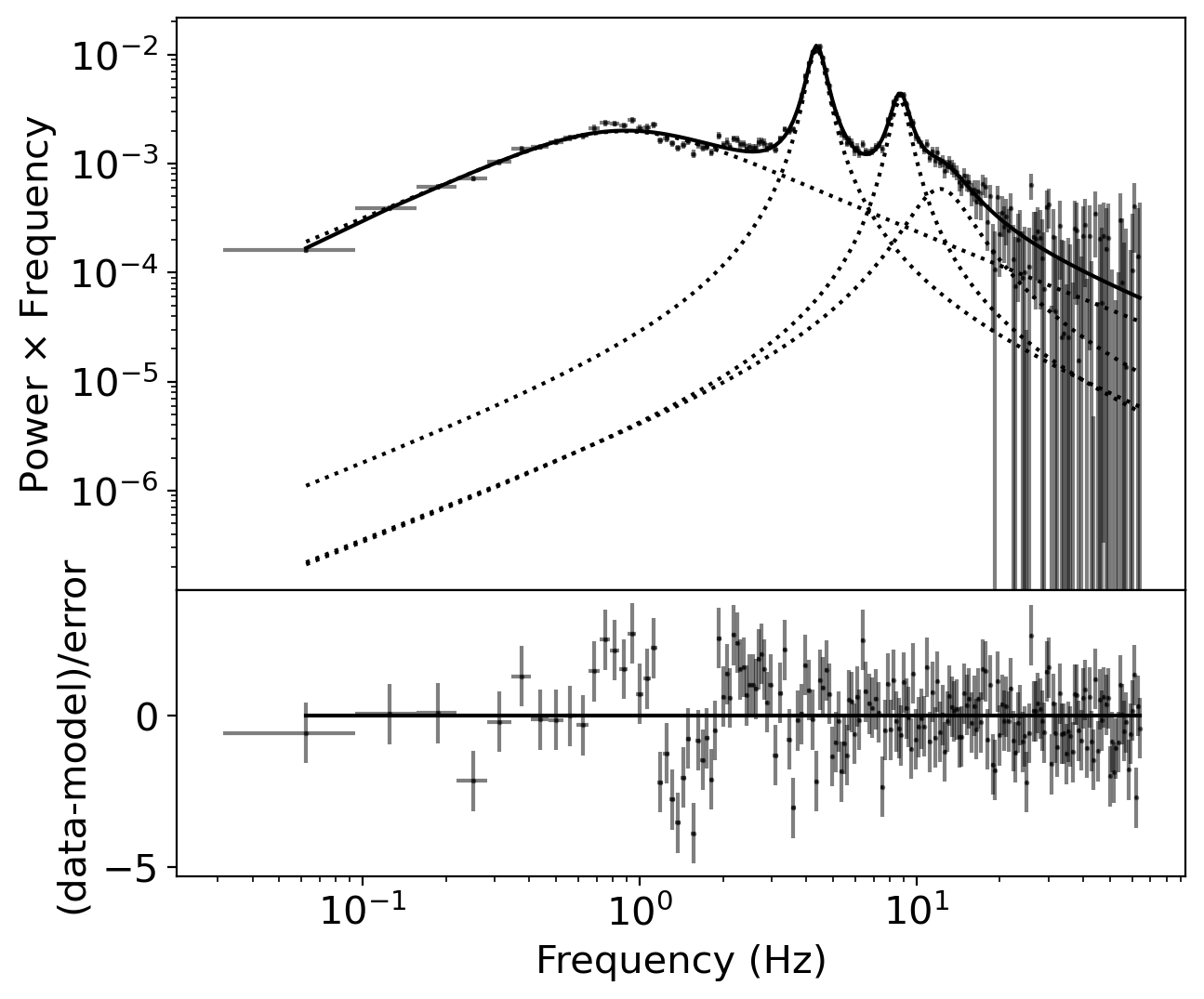}
      \end{minipage}
    }
    \caption{The bicoherence patterns and PDS when the type-C QPO appeared again after the period of type-B QPO. During this period, the bicoherence patterns are 'hypotenuse' patterns, that is, the high bicoherence is seen where two frequency components add to the frequency of QPO. The PDS is fitted with four or five Lorentzians, followed by a residual plot below.}
    \label{per2}
\end{figure*}

After a period of type-B QPOs, type-C QPOs reappear again at around MJD 58026. As shown in Fig. \ref{per2}, a diagonal line of high bicoherence appears in the patterns where $f_1+f_2=f_{QPO}$ during this period, called 'hypotenuse' patterns. We can also see high bicoherence at $f_1=f_2=f_{QPO}$, indicating the appearance of the second harmonic and the phase coupling between QPO and the second harmonic. The PDS is also presented and fitted with four or five Lorentzians. Compared with Fig. \ref{per1}, bicoherence in this stage is generally lower, maybe caused by the relatively stronger noise diluting the correlations. In this state, components of the PDS are generally different from the previous state. PDS generally has one or two more broadband noise components and the distributions of the components are also different, which might be the reason for the weakening of the vertical and horizontal streaks.

We can tell from the 'hypotenuse' and '\textbf{web}' patterns presented in \cite{Arur_2019} and \cite{Arur2020MNRAS.491..313A} that their bicoherence values along $f_1+f_2=f_{QPO}$ are generally uniform to form a continuous diagonal line. However, in our patterns, the bicoherence values along $f_1+f_2=f_{QPO}$ are higher where one frequency component approaches 0 than those where $f_1=f_2=\frac{1}{2}f_{QPO}$, so that in some patterns the diagonal line disconnects in the middle. This feature might indicate that sub-harmonic in this system is either very weak or non-existent.

\begin{figure}
    \centering
    \includegraphics[width=0.48\textwidth]{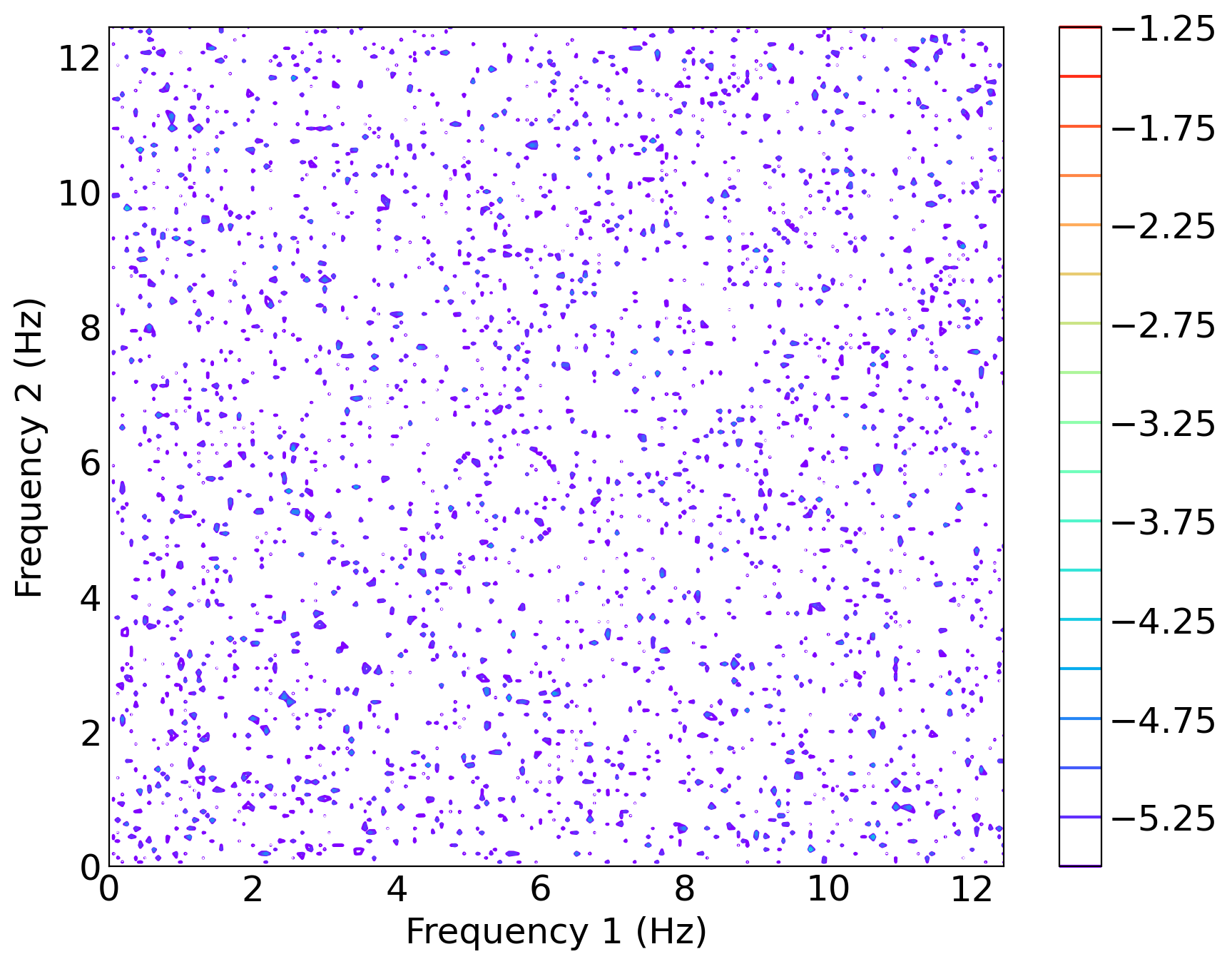}
    \caption{Bicoherence result in 1050360120 when type-B QPO is shown. No discernible pattern is shown in the plot.}
    \label{fig:bico-typeB}
\end{figure}

Throughout the outburst with type-C QPOs of MAXI J1535-571, we notice the 'web' pattern and 'hypotenuse' pattern, but the 'cross' pattern has not been discovered in this system. In addition, when type-B QPO occurs, no discernible pattern is observed in bicoherence, as shown in Fig. \ref{fig:bico-typeB}. According to \cite{Arur2020MNRAS.491..313A}, type-B QPO exhibits phase coupling exclusively with harmonics. As illustrated in Fig. \ref{qpo}, the second harmonic is barely seen, providing a plausible explanation for our results.

\section{Discussion}
\label{discuss}

\subsection{Patterns and inclination}
The change of bicoherence pattern of type-C QPOs has been reported \citep[e.g.][]{Arur_2019, Arur2020MNRAS.491..313A}. They find that the pattern transition has an inclination dependence. For low inclination sources, the pattern shifts from a 'web' to a 'hypotenuse' during the softening of the outburst, while for high inclination sources, the pattern shifts from a 'web' to a 'cross'. Note that \cite{Arur2020MNRAS.491..313A} defines sources with inclination above $\sim 65-70^{\circ}$ as high inclination sources, those below as low inclination sources, and those roughly in between as intermediate inclination sources. Despite limited data, their research indicates that intermediate inclination sources exhibit pattern changes similar to high inclination sources, i.e. from the 'web' to 'cross' pattern.

In the case of MAXI J1535-571, our bicoherence results indicate a transition from the 'web' pattern to the 'hypotenuse' pattern, implying a low inclination source. However, spectral fitting results from \cite{Miller2018ApJ...860L..28M} suggest an inclination of 67.4(8)$^{\circ}$, classifying it as an intermediate inclination source, which contradicts the bicoherence pattern transitions reported by \cite{Arur2020MNRAS.491..313A}. On the contrary, an inclination of $57_{-2}^{+1 \circ}$ derived from \cite{Xu2018ApJ...852L..34X} aligns more closely with the bicoherence-based conclusion of \cite{Arur2020MNRAS.491..313A}. Despite the possibility of disk warping or misalignment between the disk and the radio jet, \cite{Russell2019ApJ...883..198R} suggests that the jet inclination is $\leqslant 45^{\circ}$. In summary, the inclination estimates for MAXI J1535-571 exhibit discrepancies, with current bicoherence patterns suggesting a low inclination source. However, the behavior of intermediate inclination sources requires further investigation.

\subsection{Energy dependence}
\label{energy}

\begin{figure*}
    \centering
    \begin{minipage}[]{0.45\linewidth}
     \includegraphics[width=\textwidth]{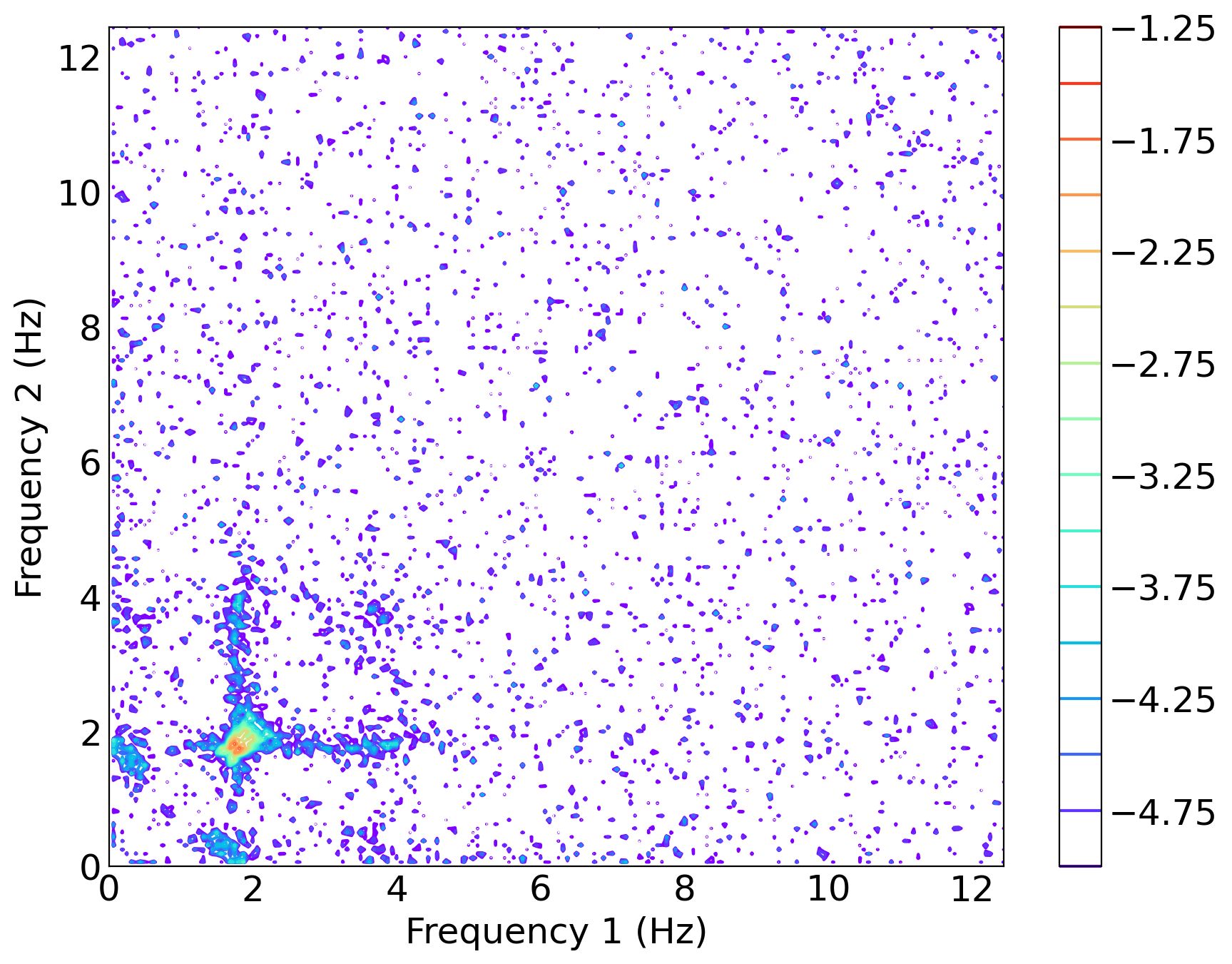}
     \flushleft
     \includegraphics[width=0.9\textwidth]{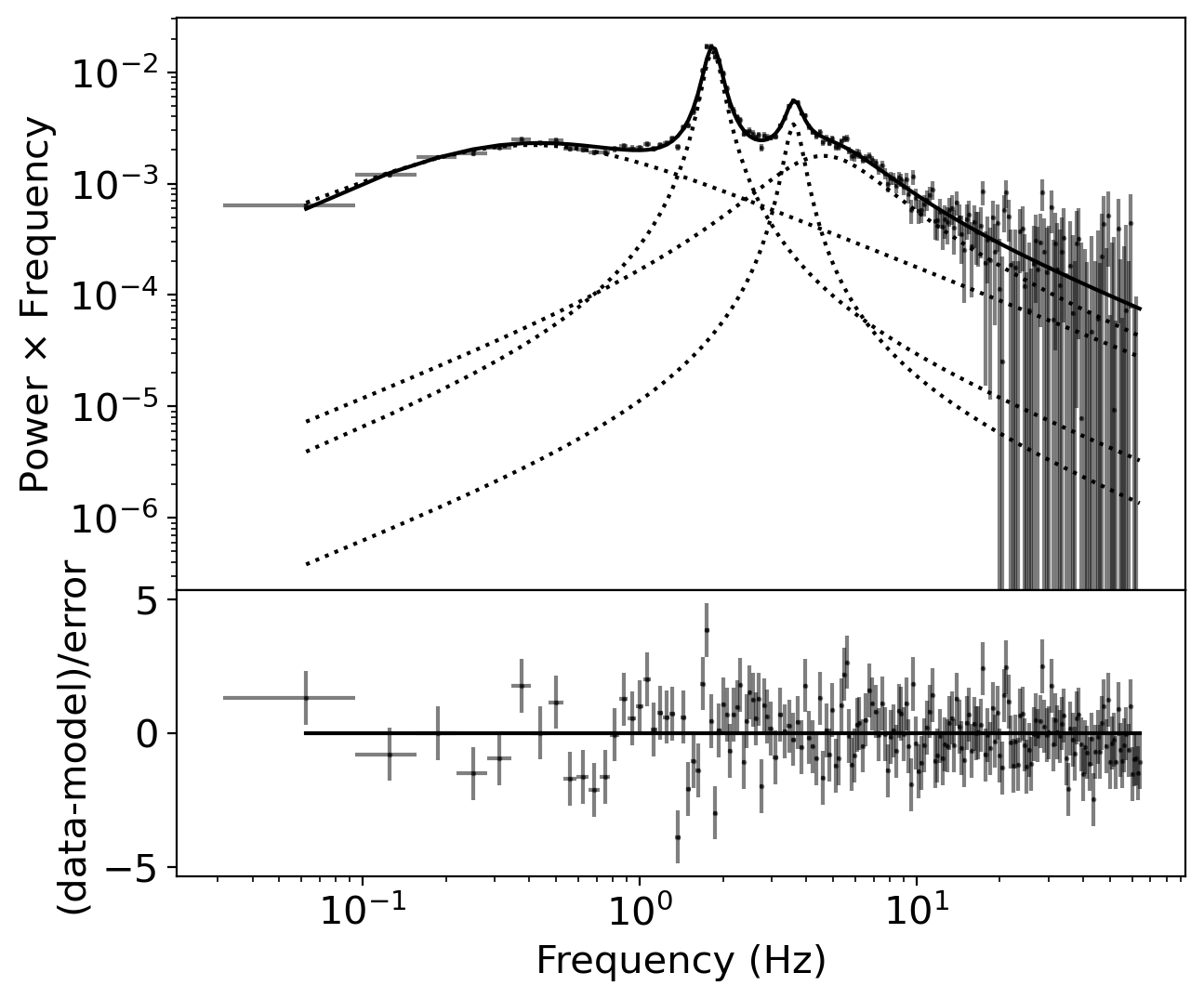}
      \end{minipage}
    \begin{minipage}[]{0.45\linewidth}
     \includegraphics[width=\textwidth]{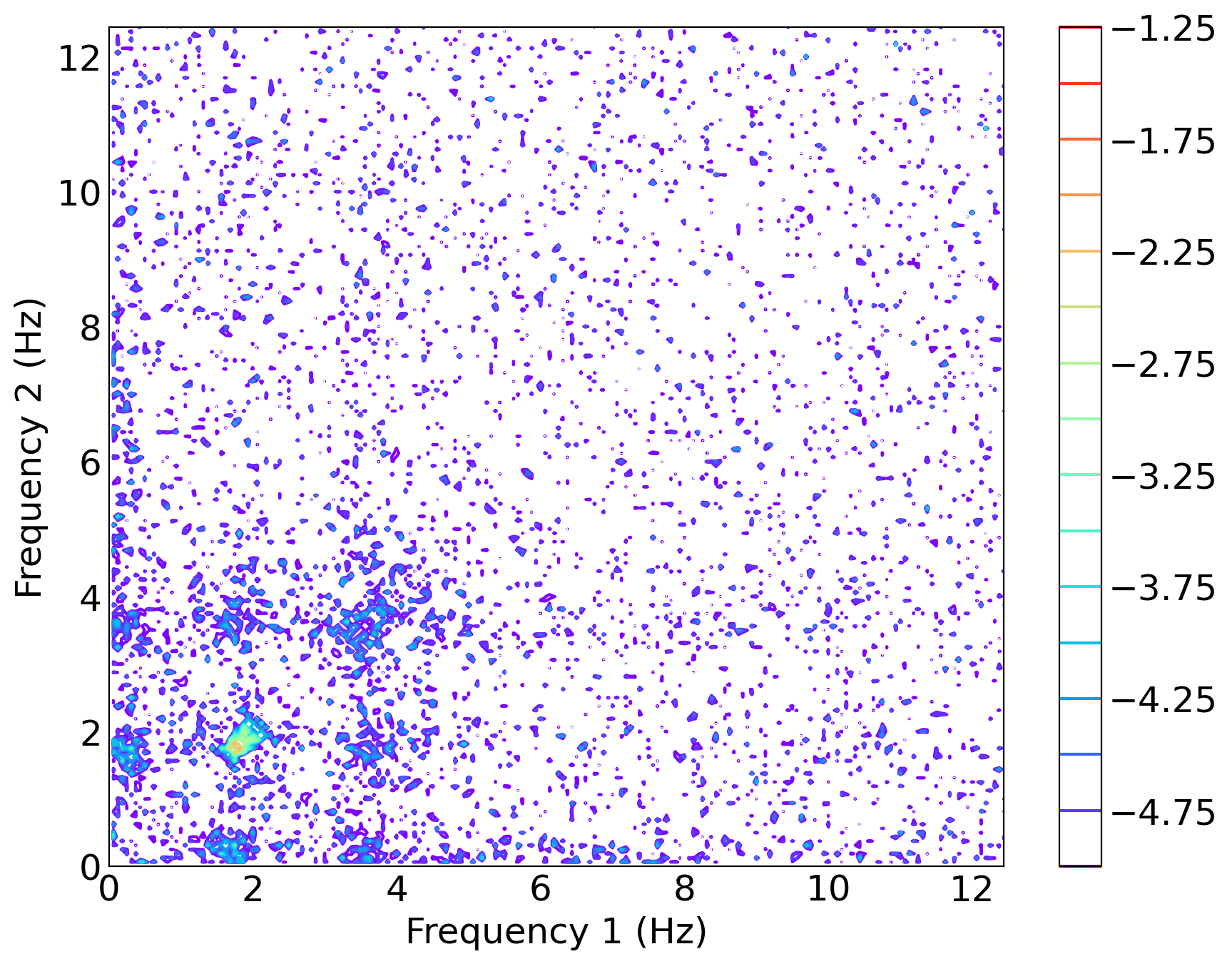}
     \flushleft
     \includegraphics[width=0.9\textwidth]{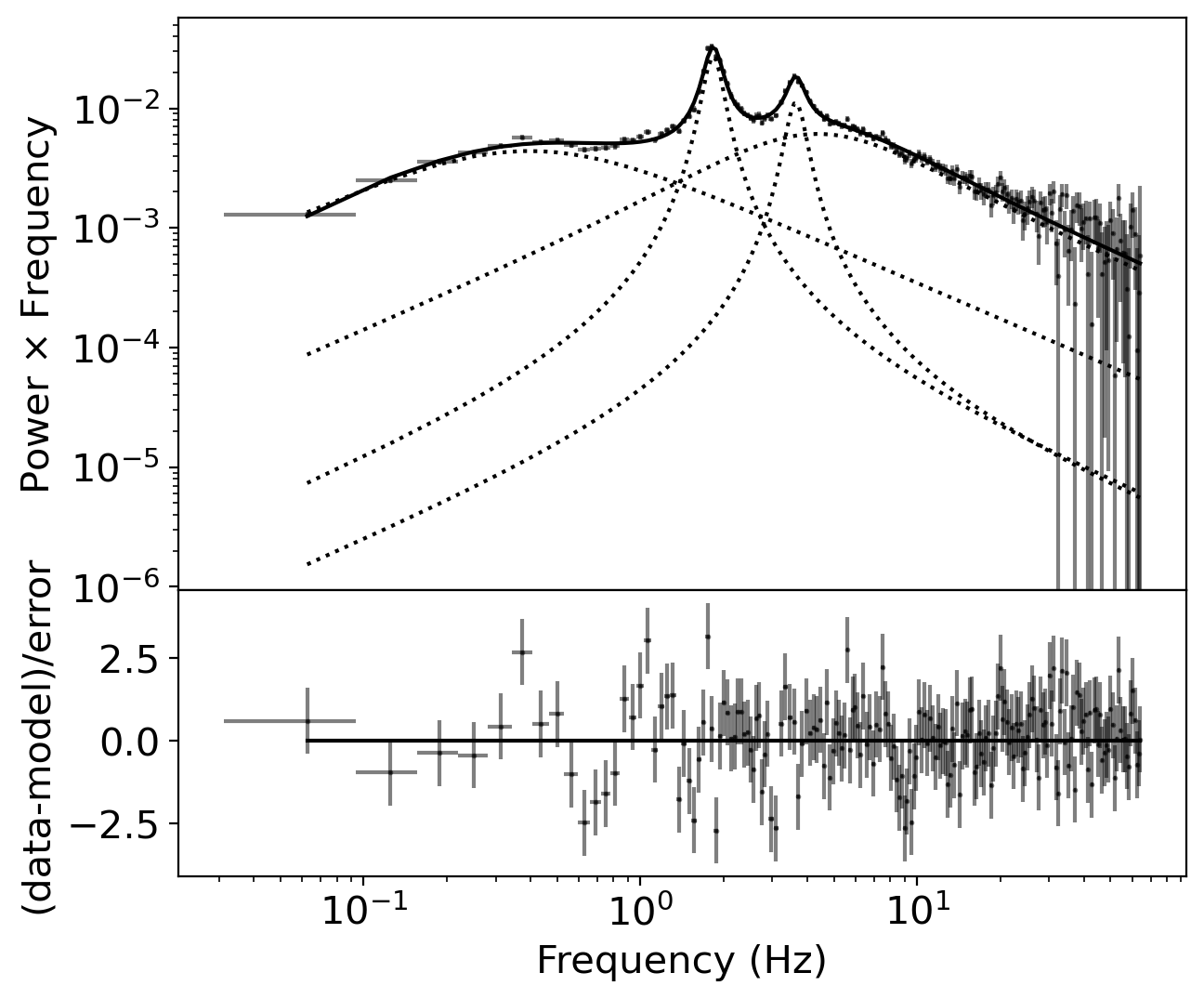}
      \end{minipage}
    \caption{Bicoherence patterns and PDS in 1-3 keV (left panels) and 3-10 keV (right panels) of Obs no.5.}
    \label{ener1}
\end{figure*}

\begin{figure*}
    \centering
        \begin{minipage}[ ]{0.45\linewidth}
        \centering
        \includegraphics[width=1\textwidth]{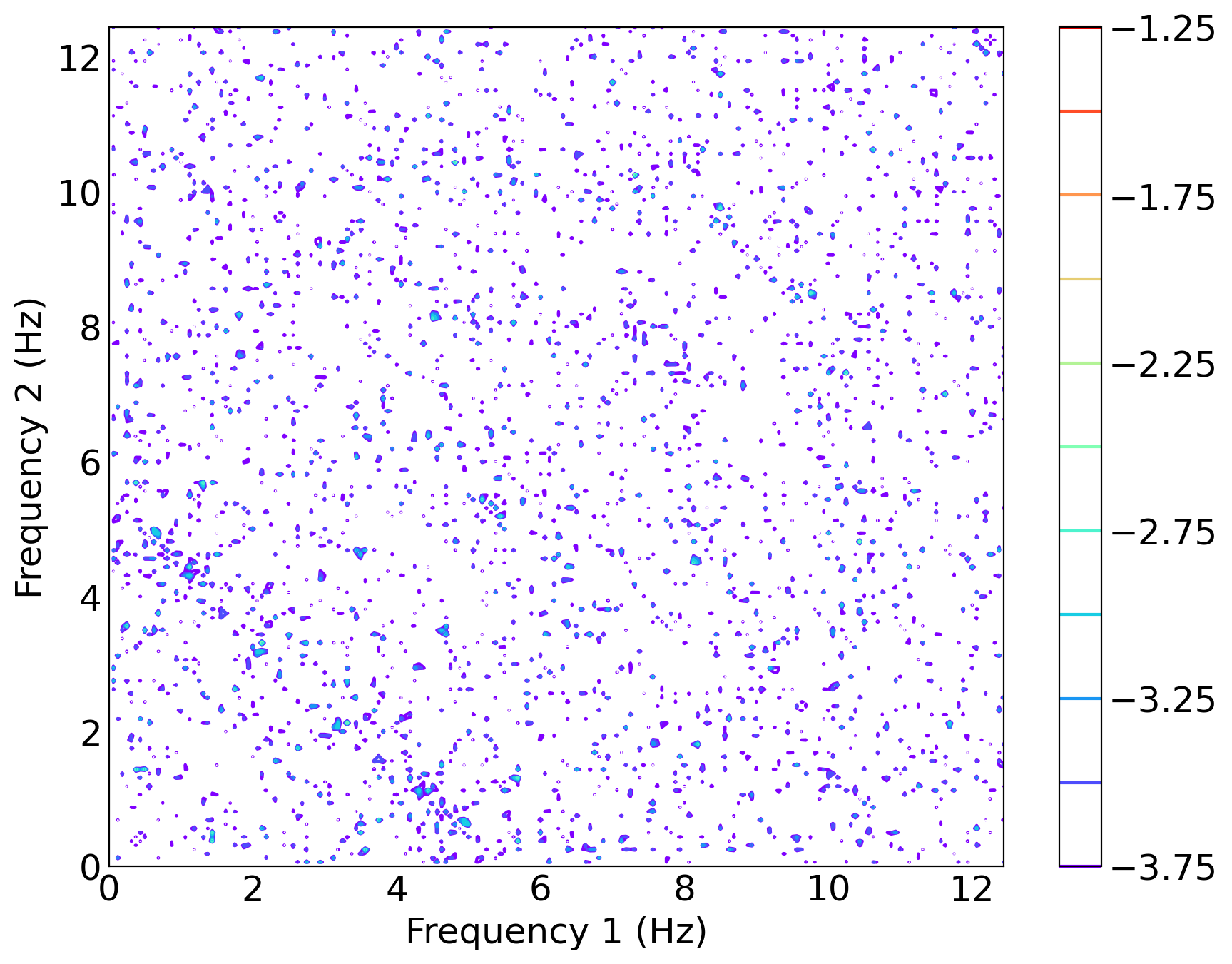}
        \flushleft
        \includegraphics[width=0.9\textwidth]{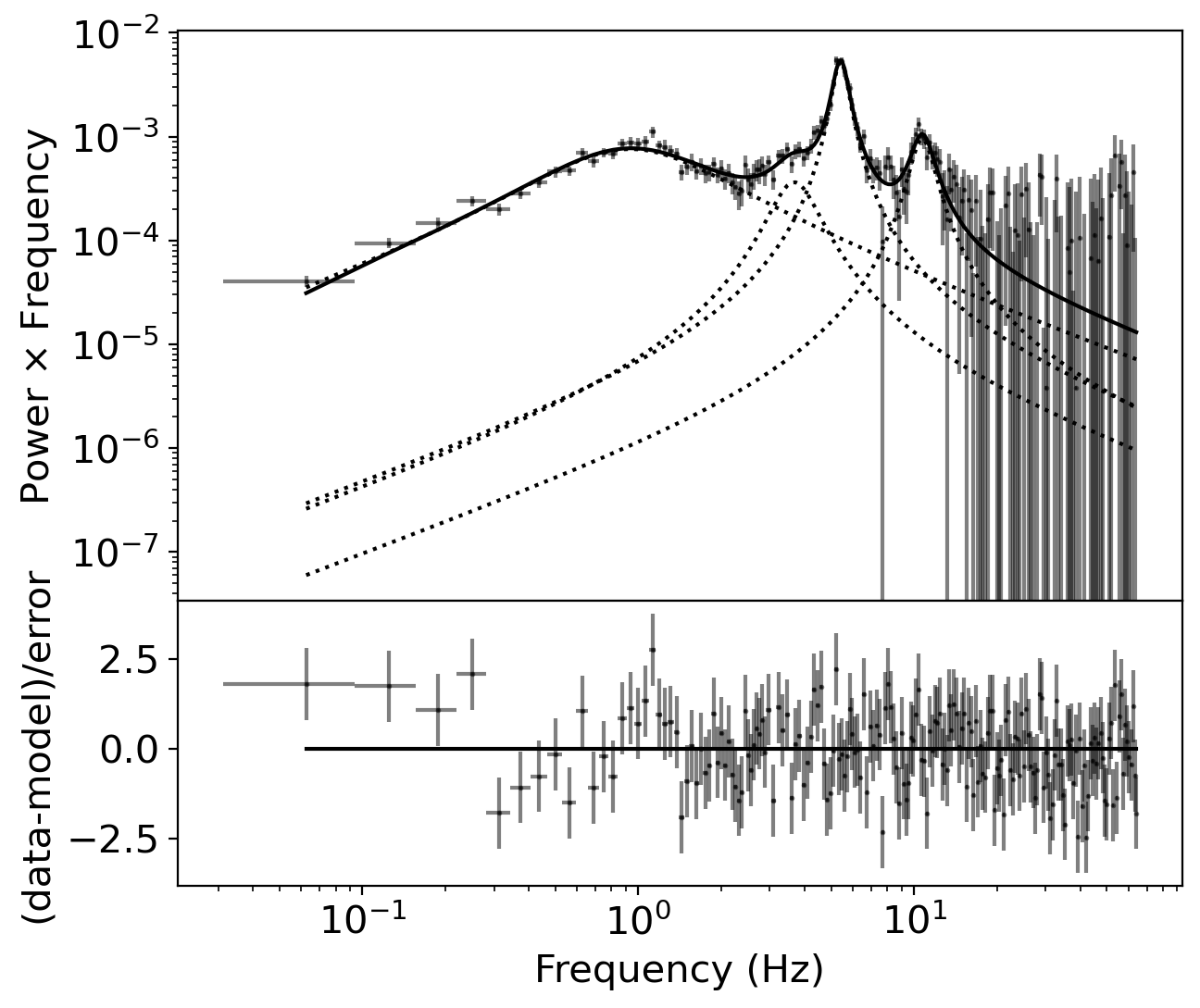}
        \end{minipage}
        \begin{minipage}[ ]{0.45\linewidth}
        \centering
        \includegraphics[width=1\textwidth]{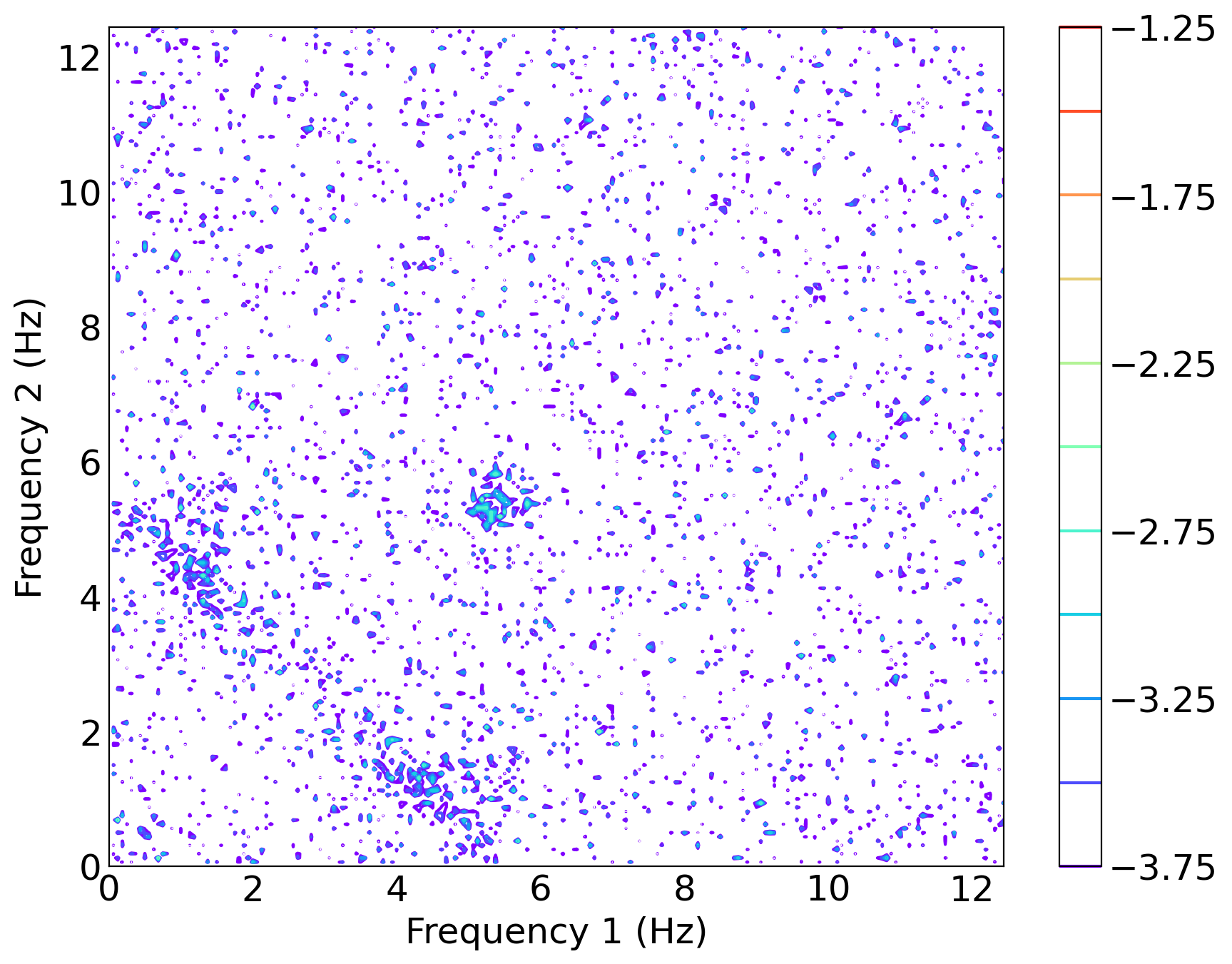}
        \flushleft
        \includegraphics[width=0.9\textwidth]{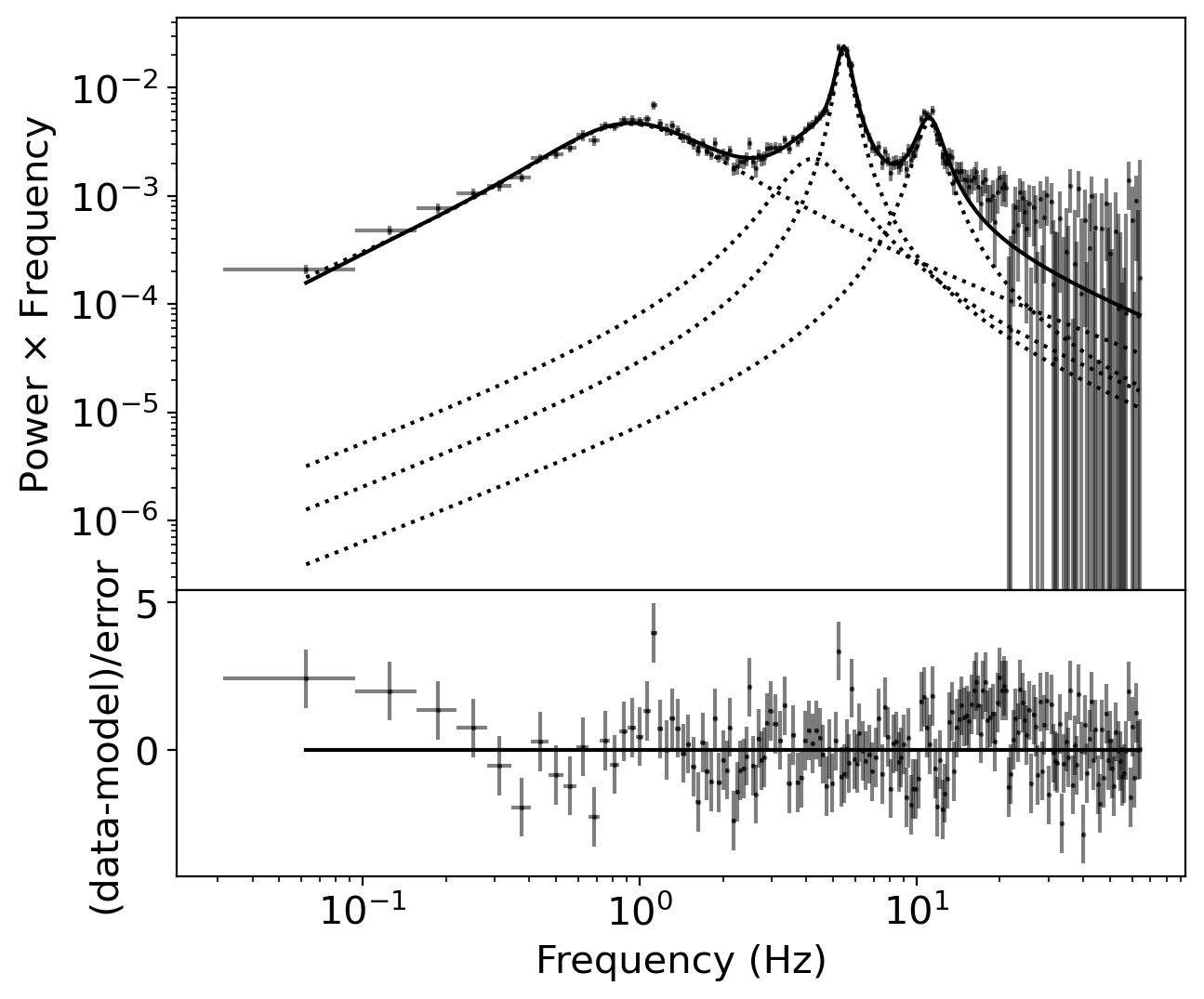}
        \end{minipage}
    \caption{Bicoherence patterns and PDS in 1-3 keV (left panels) and 3-10 keV (right panels) of Obs no.14.}
    \label{ener2}
\end{figure*}

To investigate the energy dependence of patterns, we compare bicoherence patterns in two energy bands: 1-3 and 3-10 keV. During the period when type-C QPO first appeared, bicoherence patterns in 1-3 keV generally have stronger horizontal and vertical lines, showing strong coupling between the QPO fundamental and the noise components, while bicoherence patterns in 3-10 keV show stronger phase coupling between QPO fundamental and harmonics. Here, we take the results of different energy bands of obs no. 5 as an example shown in Fig. \ref{ener1}, with their PDS presented in the bottom. The PDS does not show significant differences for different energy bands. However, it should be noted that the count rates in 1-3 keV ($\sim 4100 cts/s$) are generally higher than those in 3-10 keV ($\sim 2600 cts/s$), so we can not tell whether the stronger horizontal and vertical lines are really due to different physical attributes in two bands or affected by the higher count rates. After type-C QPO appears again, we notice that bicoherence patterns in 3-10 keV are generally more evident than in 1-3 keV. See Fig. \ref{ener2} as an example. Given the higher count rates in the 1-3 keV range, we are more confident in attributing this result to a physical cause. We have not found any pattern differences of different energy bands when the type-B QPO appears.

\subsection{Corona evolution}

As suggested by \cite{Arur2020MNRAS.491..313A}, the 'web' pattern is a combination of the diagonal feature in the 'hypotenuse' and the vertical and horizontal feature (hereafter cross feature) in the 'cross' pattern. The diagonal feature of high bicoherence is produced by the upscattering of thin disc photons within the corona, while the cross feature is related to the precession of the inner accretion flow originating from the inner regions of the disc. This implies that the 'web' pattern is simultaneously affected by two corona components. As \cite{Rawat2023MNRAS.520..113R} pointed out, a dual corona structure may indeed exist for MAXI J1535$-$571. From now on, we refer to the corona responsible for generating the cross feature as Corona 1, and the one producing the diagonal feature as Corona 2.

\begin{figure*}
    \includegraphics[width=0.85\textwidth,height=0.85\columnwidth]{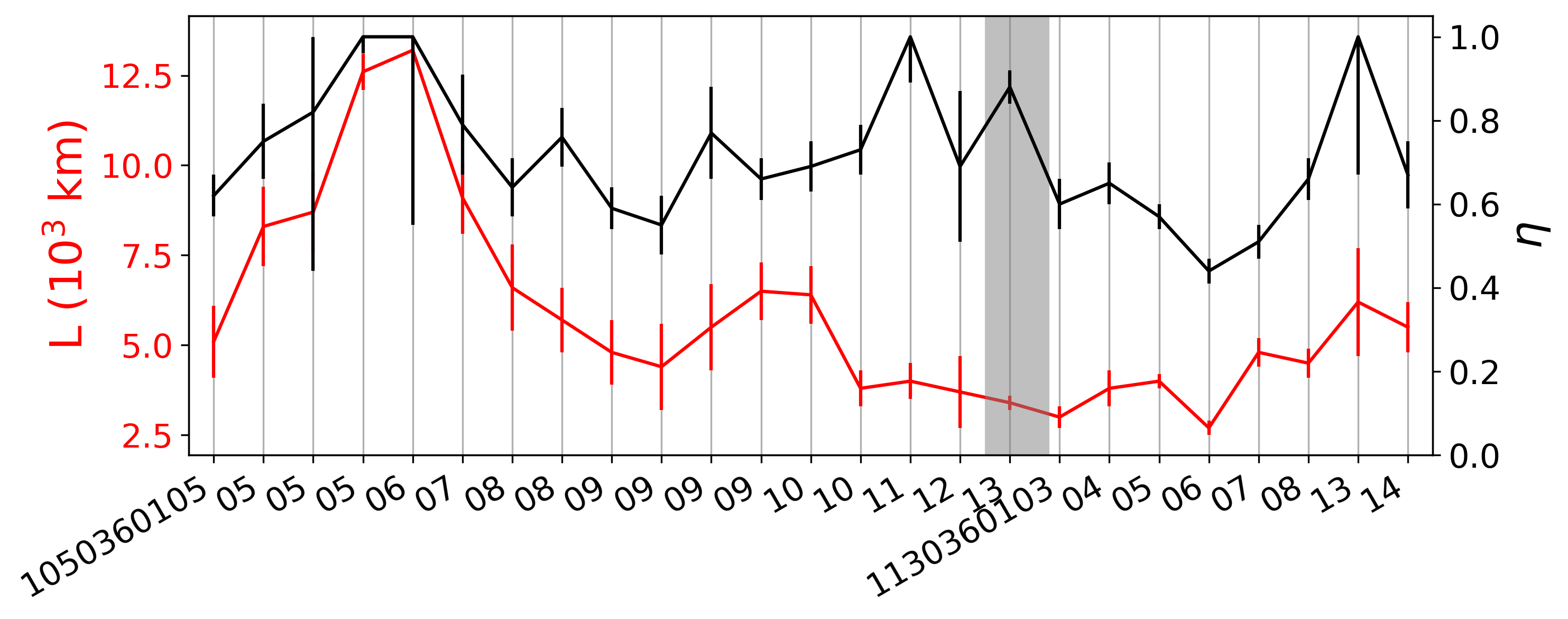}
    \caption{The corona size ($L$, the red line) and the feedback fraction ($\eta$, the black line) evolution. X-axis marks the NICER observation IDs. Two-digit IDs indicate that the preceding digits are identical to those on the left. The grey area roughly outlines the SIMS.}
    \label{fig:test}
\end{figure*}

In Figure~\ref{fig:test} we plot the size of the corona, $L$, and the feedback fraction, $\eta$, obtained from Table 2 of \cite{Rawat2023MNRAS.520..113R}. Comparing with our bicoherence results in Figure~\ref{per1}, we find that the cross pattern is most distinct when both corona size and feedback fraction reach their maximum (obs no. 5, i.e. ObsID 1050360106) and gradually diminishes as the two parameters decrease (obs no. 6, i.e. ObsID 1050360107). This suggests that Corona 1 is gradually diminishing.

\begin{figure}
    \centering
        \begin{minipage}[Obs no.12]{\linewidth}
        \centering
        \includegraphics[width=1\textwidth]{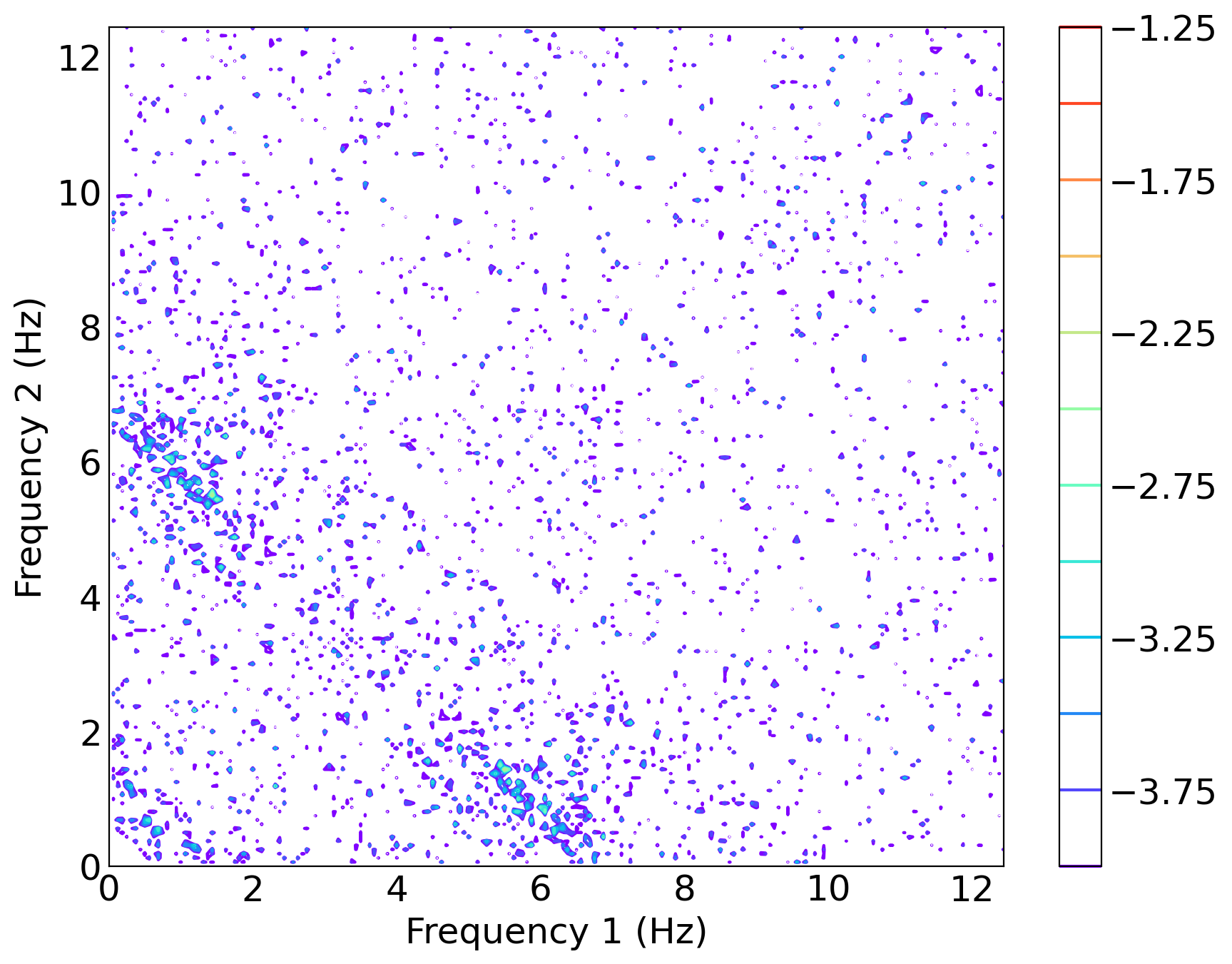}
        \flushleft
        \includegraphics[width=0.9\textwidth]{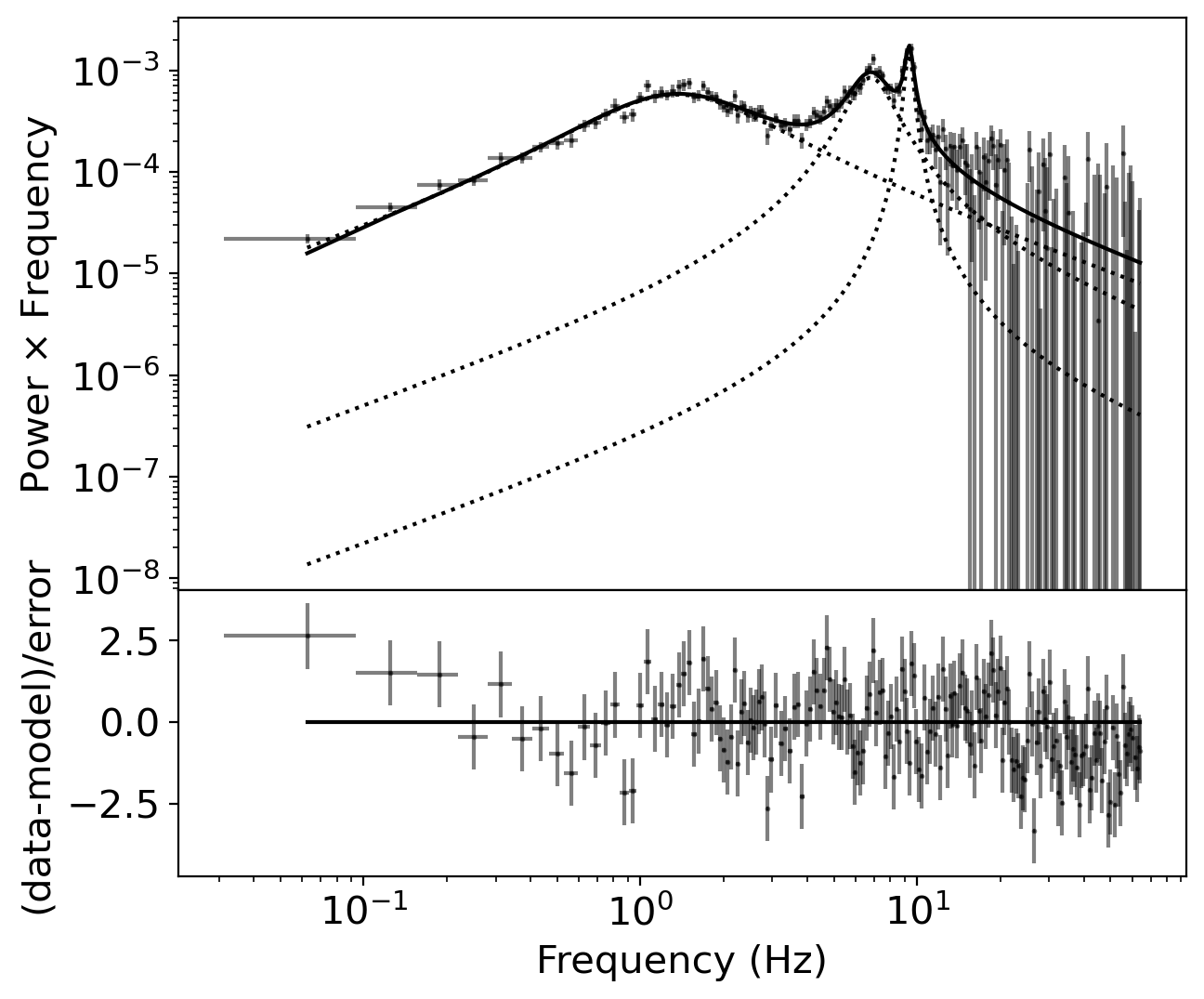}
        \end{minipage}
	\caption{The bicoherence pattern and corresponding PDS of obs no.12 in the early stage of the SIMS.}
	\label{fig:obs12}
\end{figure}

Subsequently, the hypotenuse pattern becomes more evident during the transient expansion of the corona in ObsID 1050360109 (obs no. 7), and is still visible when the system just enters the SIMS (obs no. 12, i.e. ObsID 1050360113, see the example in Figure~\ref{fig:obs12}), even though it is very weak. This short expansion of the corona with relatively small feedback fraction is possibly related to jets  \citep{Rawat2023MNRAS.520..113R,Zhang_2022,Zhang2023MNRAS.520.5144Z}. After the source fully enters the SIMS, our analysis does not reveal any discernible pattern during the occurrence of type-B QPOs, not even for the second harmonic feature as indicated by \cite{Arur2020MNRAS.491..313A}. Therefore, we hypothesize that there is a certain connection between the hypotenuse pattern and the jets. The jet in MAXI J1535-571 may take away a portion of the corona, leading to the emergence of low bicoherence in our SIMS. In other words, during this stage, Corona 2 increases and gradually forms a jet, then it decreases again as the jet is emitted.

When the source re-enters the HIMS, the hypotenuse pattern reappears (obs no. 14, see panel (a) in Figure~\ref{per2} and ObsID 1130360104 in Figure~\ref{fig:test}), indicating the replenishment of Corona 2. Interestingly, at obs no. 20 (see panel (c) in Figure~\ref{per2}, and ObsID 1130360113 in Figure~\ref{fig:test}), the size of corona is at its maximum after SIMS with a relatively high $\eta$, yet the hypotenuse pattern is not prominent, even though the error bar is large. Instead, it becomes evident when both $\eta$ and $L$ decrease (obs no. 21, i.e. ObsID 1130360114 in Figure~\ref{fig:test}), suggesting an increase in Corona 1 during the final stage, which may suppress the signals from Corona 2.

\section{Summary}
\label{summ}
In this study, we investigated the relationship between QPO types and patterns in MAXI J1535-571 using bicoherence. We have found that type-C QPO in MAXI J1535-571 is coupled with higher harmonics and broadband noise. In addition, the bicoherence is a 'web' pattern when the type-C QPO first appears and becomes a 'hypotenuse' pattern when the type-C QPO appears again after the period of the type-B QPO. However, we do not find any patterns during the occurrence of type-B QPOs. In addition, our results reveal differences in bicoherence intensity between 1-3 keV and 3-10 keV, showing that the intensity of bicoherence may be correlated with energy bands. The horizontal and vertical lines in the 'web' pattern are stronger in lower energy bands before type-B QPO appears, and diagonal lines in the 'hypotenuse' pattern are stronger in higher energy bands after type-B QPO appears. Furthermore, based on the inclination dependence proposed by \cite{Arur2020MNRAS.491..313A}, we constrain the inclination of MAXI J1535-571 which would be a low inclination source. Finally, we discuss the corona evolution process using dual-corona geometry during the state transitions.

\section*{Acknowledgements}
This work is supported by the NSFC (Grants No. 12133007) and the National Key Research and Development Program of China (Grants No. 2021YFA0718503).
\section*{Data Availability}
The data of \textit{NICER} used in this article can be obtained from this website \url{https://heasarc.gsfc.nasa.gov/cgi-bin/W3Browse/w3browse.pl}.

\bibliographystyle{mnras}
\bibliography{example}

\begin{thebibliography}{}
\makeatletter
\relax
\def\mn@urlcharsother{\let\do\@makeother \do\$\do\&\do\#\do\^\do\_\do\%\do\~}
\def\mn@doi{\begingroup\mn@urlcharsother \@ifnextchar [ {\mn@doi@}
  {\mn@doi@[]}}
\def\mn@doi@[#1]#2{\def\@tempa{#1}\ifx\@tempa\@empty \href
  {http://dx.doi.org/#2} {doi:#2}\else \href {http://dx.doi.org/#2} {#1}\fi
  \endgroup}
\def\mn@eprint#1#2{\mn@eprint@#1:#2::\@nil}
\def\mn@eprint@arXiv#1{\href {http://arxiv.org/abs/#1} {{\tt arXiv:#1}}}
\def\mn@eprint@dblp#1{\href {http://dblp.uni-trier.de/rec/bibtex/#1.xml}
  {dblp:#1}}
\def\mn@eprint@#1:#2:#3:#4\@nil{\def\@tempa {#1}\def\@tempb {#2}\def\@tempc
  {#3}\ifx \@tempc \@empty \let \@tempc \@tempb \let \@tempb \@tempa \fi \ifx
  \@tempb \@empty \def\@tempb {arXiv}\fi \@ifundefined
  {mn@eprint@\@tempb}{\@tempb:\@tempc}{\expandafter \expandafter \csname
  mn@eprint@\@tempb\endcsname \expandafter{\@tempc}}}

\bibitem[\protect\citeauthoryear{Arur \& Maccarone}{Arur \&
  Maccarone}{2019}]{Arur_2019}
Arur K.,  Maccarone T.~J.,  2019, \mn@doi [Monthly Notices of the Royal
  Astronomical Society] {10.1093/mnras/stz1052}

\bibitem[\protect\citeauthoryear{{Arur} \& {Maccarone}}{{Arur} \&
  {Maccarone}}{2020}]{Arur2020MNRAS.491..313A}
{Arur} K.,  {Maccarone} T.~J.,  2020, \mn@doi [\mnras] {10.1093/mnras/stz2784},
  \href {https://ui.adsabs.harvard.edu/abs/2020MNRAS.491..313A} {491, 313}

\bibitem[\protect\citeauthoryear{{Belloni}, {Psaltis}  \& {van der
  Klis}}{{Belloni} et~al.}{2002}]{Belloni2002ApJ...572..392B}
{Belloni} T.,  {Psaltis} D.,   {van der Klis} M.,  2002, \mn@doi [\apj]
  {10.1086/340290}, \href
  {https://ui.adsabs.harvard.edu/abs/2002ApJ...572..392B} {572, 392}

\bibitem[\protect\citeauthoryear{{Bhargava}, {Belloni}, {Bhattacharya}  \&
  {Misra}}{{Bhargava} et~al.}{2019}]{Bhargava2019MNRAS.488..720B}
{Bhargava} Y.,  {Belloni} T.,  {Bhattacharya} D.,   {Misra} R.,  2019, \mn@doi
  [\mnras] {10.1093/mnras/stz1774}, \href
  {https://ui.adsabs.harvard.edu/abs/2019MNRAS.488..720B} {488, 720}

\bibitem[\protect\citeauthoryear{{Casella}, {Belloni}  \& {Stella}}{{Casella}
  et~al.}{2005}]{Casella2005ApJ...629..403C}
{Casella} P.,  {Belloni} T.,   {Stella} L.,  2005, \mn@doi [\apj]
  {10.1086/431174}, \href
  {https://ui.adsabs.harvard.edu/abs/2005ApJ...629..403C} {629, 403}

\bibitem[\protect\citeauthoryear{Fender, Belloni  \& Gallo}{Fender
  et~al.}{2004}]{10.1111/j.1365-2966.2004.08384.x}
Fender R.~P.,  Belloni T.~M.,   Gallo E.,  2004, \mn@doi [Monthly Notices of
  the Royal Astronomical Society] {10.1111/j.1365-2966.2004.08384.x}, 355, 1105

\bibitem[\protect\citeauthoryear{{Fender}, {Homan}  \& {Belloni}}{{Fender}
  et~al.}{2009}]{Fender2009MNRAS.396.1370F}
{Fender} R.~P.,  {Homan} J.,   {Belloni} T.~M.,  2009, \mn@doi [\mnras]
  {10.1111/j.1365-2966.2009.14841.x}, \href
  {https://ui.adsabs.harvard.edu/abs/2009MNRAS.396.1370F} {396, 1370}

\bibitem[\protect\citeauthoryear{{Gendreau} et~al.,}{{Gendreau}
  et~al.}{2017}]{Gendreau2017ATel10768....1G}
{Gendreau} K.,  et~al., 2017, The Astronomer's Telegram, \href
  {https://ui.adsabs.harvard.edu/abs/2017ATel10768....1G} {10768, 1}

\bibitem[\protect\citeauthoryear{{Hjellming} \& {Rupen}}{{Hjellming} \&
  {Rupen}}{1995}]{Hjellming1995Natur.375..464H}
{Hjellming} R.~M.,  {Rupen} M.~P.,  1995, \mn@doi [\nat] {10.1038/375464a0},
  \href {https://ui.adsabs.harvard.edu/abs/1995Natur.375..464H} {375, 464}

\bibitem[\protect\citeauthoryear{Homan, Wijnands, van~der Klis, Belloni, van
  Paradijs, Klein-Wolt, Fender  \& Mendez}{Homan et~al.}{2001}]{Homan_2001}
Homan J.,  Wijnands R.,  van~der Klis M.,  Belloni T.,  van Paradijs J.,
  Klein-Wolt M.,  Fender R.,   Mendez M.,  2001, \mn@doi [The Astrophysical
  Journal Supplement Series] {10.1086/318954}, 132, 377

\bibitem[\protect\citeauthoryear{{Huang} et~al.,}{{Huang}
  et~al.}{2018}]{Huang2018ApJ...866..122H}
{Huang} Y.,  et~al., 2018, \mn@doi [\apj] {10.3847/1538-4357/aade4c}, \href
  {https://ui.adsabs.harvard.edu/abs/2018ApJ...866..122H} {866, 122}

\bibitem[\protect\citeauthoryear{Ingram, Done  \& Fragile}{Ingram
  et~al.}{2009}]{Ingram.2009.00693.x}
Ingram A.,  Done C.,   Fragile P.~C.,  2009, \mn@doi [Monthly Notices of the
  Royal Astronomical Society: Letters] {10.1111/j.1745-3933.2009.00693.x}, 397,
  L101

\bibitem[\protect\citeauthoryear{{Kennea}}{{Kennea}}{2017}]{Kenneaalone2017ATel10731....1K}
{Kennea} J.~A.,  2017, The Astronomer's Telegram, \href
  {https://ui.adsabs.harvard.edu/abs/2017ATel10731....1K} {10731, 1}

\bibitem[\protect\citeauthoryear{{Kennea}, {Evans}, {Beardmore}, {Krimm},
  {Romano}, {Yamaoka}, {Serino}  \& {Negoro}}{{Kennea}
  et~al.}{2017}]{Kennea2017ATel10700....1K}
{Kennea} J.~A.,  {Evans} P.~A.,  {Beardmore} A.~P.,  {Krimm} H.~A.,  {Romano}
  P.,  {Yamaoka} K.,  {Serino} M.,   {Negoro} H.,  2017, The Astronomer's
  Telegram, \href {https://ui.adsabs.harvard.edu/abs/2017ATel10700....1K}
  {10700, 1}

\bibitem[\protect\citeauthoryear{Kim \& Powers}{Kim \& Powers}{1979}]{4317207}
Kim Y.~C.,  Powers E.~J.,  1979, \mn@doi [IEEE Transactions on Plasma Science]
  {10.1109/TPS.1979.4317207}, 7, 120

\bibitem[\protect\citeauthoryear{Maccarone \& Coppi}{Maccarone \&
  Coppi}{2003}]{Maccarone.2003.06040.x}
Maccarone T.~J.,  Coppi P.~S.,  2003, \mn@doi [Monthly Notices of the Royal
  Astronomical Society] {10.1046/j.1365-8711.2003.06040.x}, 338, 189

\bibitem[\protect\citeauthoryear{Maccarone \& Schnittman}{Maccarone \&
  Schnittman}{2005}]{Maccarone2004.08615.x}
Maccarone T.~J.,  Schnittman J.~D.,  2005, \mn@doi [Monthly Notices of the
  Royal Astronomical Society] {10.1111/j.1365-2966.2004.08615.x}, 357, 12

\bibitem[\protect\citeauthoryear{Maccarone, Uttley, van~der Klis, Wijnands  \&
  Coppi}{Maccarone et~al.}{2011}]{Maccarone_2011}
Maccarone T.~J.,  Uttley P.,  van~der Klis M.,  Wijnands R. A.~D.,   Coppi
  P.~S.,  2011, \mn@doi [Monthly Notices of the Royal Astronomical Society]
  {10.1111/j.1365-2966.2011.18273.x}, 413, 1819

\bibitem[\protect\citeauthoryear{{Mereminskiy} \& {Grebenev}}{{Mereminskiy} \&
  {Grebenev}}{2017}]{Mereminskiy2017ATel10734....1M}
{Mereminskiy} I.~A.,  {Grebenev} S.~A.,  2017, The Astronomer's Telegram, \href
  {https://ui.adsabs.harvard.edu/abs/2017ATel10734....1M} {10734, 1}

\bibitem[\protect\citeauthoryear{{Miller-Jones} et~al.,}{{Miller-Jones}
  et~al.}{2012}]{Miller-Jones2012MNRAS.421..468M}
{Miller-Jones} J.~C.~A.,  et~al., 2012, \mn@doi [\mnras]
  {10.1111/j.1365-2966.2011.20326.x}, \href
  {https://ui.adsabs.harvard.edu/abs/2012MNRAS.421..468M} {421, 468}

\bibitem[\protect\citeauthoryear{{Miller-Jones} et~al.,}{{Miller-Jones}
  et~al.}{2019}]{Miller-Jones2019Natur.569..374M}
{Miller-Jones} J. C.~A.,  et~al., 2019, \mn@doi [\nat]
  {10.1038/s41586-019-1152-0}, \href
  {https://ui.adsabs.harvard.edu/abs/2019Natur.569..374M} {569, 374}

\bibitem[\protect\citeauthoryear{{Miller} et~al.,}{{Miller}
  et~al.}{2018}]{Miller2018ApJ...860L..28M}
{Miller} J.~M.,  et~al., 2018, \mn@doi [\apjl] {10.3847/2041-8213/aacc61},
  \href {https://ui.adsabs.harvard.edu/abs/2018ApJ...860L..28M} {860, L28}

\bibitem[\protect\citeauthoryear{{Mirabel} \& {Rodr{\'\i}guez}}{{Mirabel} \&
  {Rodr{\'\i}guez}}{1994}]{Mirabel1994Natur.371...46M}
{Mirabel} I.~F.,  {Rodr{\'\i}guez} L.~F.,  1994, \mn@doi [\nat]
  {10.1038/371046a0}, \href
  {https://ui.adsabs.harvard.edu/abs/1994Natur.371...46M} {371, 46}

\bibitem[\protect\citeauthoryear{{Nakahira} et~al.,}{{Nakahira}
  et~al.}{2017}]{Nakahira2017ATel10729....1N}
{Nakahira} S.,  et~al., 2017, The Astronomer's Telegram, \href
  {https://ui.adsabs.harvard.edu/abs/2017ATel10729....1N} {10729, 1}

\bibitem[\protect\citeauthoryear{{Negoro} et~al.,}{{Negoro}
  et~al.}{2017}]{Negoro2017ATel10708....1N}
{Negoro} H.,  et~al., 2017, The Astronomer's Telegram, \href
  {https://ui.adsabs.harvard.edu/abs/2017ATel10708....1N} {10708, 1}

\bibitem[\protect\citeauthoryear{{Palmer}, {Krimm}  \& {Swift/BAT
  Team}}{{Palmer} et~al.}{2017}]{Palmer2017ATel10733....1P}
{Palmer} D.~M.,  {Krimm} H.~A.,   {Swift/BAT Team} 2017, The Astronomer's
  Telegram, \href {https://ui.adsabs.harvard.edu/abs/2017ATel10733....1P}
  {10733, 1}

\bibitem[\protect\citeauthoryear{{Rawat} et~al.,}{{Rawat}
  et~al.}{2023}]{Rawat2023MNRAS.520..113R}
{Rawat} D.,  et~al., 2023, \mn@doi [\mnras] {10.1093/mnras/stad126}, \href
  {https://ui.adsabs.harvard.edu/abs/2023MNRAS.520..113R} {520, 113}

\bibitem[\protect\citeauthoryear{{Russell} et~al.,}{{Russell}
  et~al.}{2019}]{Russell2019ApJ...883..198R}
{Russell} T.~D.,  et~al., 2019, \mn@doi [\apj] {10.3847/1538-4357/ab3d36},
  \href {https://ui.adsabs.harvard.edu/abs/2019ApJ...883..198R} {883, 198}

\bibitem[\protect\citeauthoryear{{Shidatsu} et~al.,}{{Shidatsu}
  et~al.}{2017a}]{Shidatsu2017ATel10761....1S}
{Shidatsu} M.,  et~al., 2017a, The Astronomer's Telegram, \href
  {https://ui.adsabs.harvard.edu/abs/2017ATel10761....1S} {10761, 1}

\bibitem[\protect\citeauthoryear{{Shidatsu} et~al.,}{{Shidatsu}
  et~al.}{2017b}]{Shidatsu2017ATel11020....1S}
{Shidatsu} M.,  et~al., 2017b, The Astronomer's Telegram, \href
  {https://ui.adsabs.harvard.edu/abs/2017ATel11020....1S} {11020, 1}

\bibitem[\protect\citeauthoryear{Stella \& Vietri}{Stella \&
  Vietri}{1997}]{Stella_1998}
Stella L.,  Vietri M.,  1997, \mn@doi [The Astrophysical Journal]
  {10.1086/311075}, 492, L59

\bibitem[\protect\citeauthoryear{{Stevens} et~al.,}{{Stevens}
  et~al.}{2018}]{Stevens2018ApJ...865L..15S}
{Stevens} A.~L.,  et~al., 2018, \mn@doi [\apjl] {10.3847/2041-8213/aae1a4},
  \href {https://ui.adsabs.harvard.edu/abs/2018ApJ...865L..15S} {865, L15}

\bibitem[\protect\citeauthoryear{{Tetarenko} et~al.,}{{Tetarenko}
  et~al.}{2017a}]{Tetarenko2017MNRAS.469.3141T}
{Tetarenko} A.~J.,  et~al., 2017a, \mn@doi [\mnras] {10.1093/mnras/stx1048},
  \href {https://ui.adsabs.harvard.edu/abs/2017MNRAS.469.3141T} {469, 3141}

\bibitem[\protect\citeauthoryear{{Tetarenko}, {Russell}, {Miller-Jones},
  {Sivakoff}  \& {Jacpot Xrb Collaboration}}{{Tetarenko}
  et~al.}{2017b}]{Tetarenko2017ATel10745....1T}
{Tetarenko} A.~J.,  {Russell} T.~D.,  {Miller-Jones} J.~C.~A.,  {Sivakoff}
  G.~R.,   {Jacpot Xrb Collaboration} 2017b, The Astronomer's Telegram, \href
  {https://ui.adsabs.harvard.edu/abs/2017ATel10745....1T} {10745, 1}

\bibitem[\protect\citeauthoryear{{Vincentelli} et~al.,}{{Vincentelli}
  et~al.}{2021}]{Vincentelli2021MNRAS.503..614V}
{Vincentelli} F.~M.,  et~al., 2021, \mn@doi [\mnras] {10.1093/mnras/stab475},
  \href {https://ui.adsabs.harvard.edu/abs/2021MNRAS.503..614V} {503, 614}

\bibitem[\protect\citeauthoryear{{Wijnands}, {Homan}  \& {van der
  Klis}}{{Wijnands} et~al.}{1999}]{Wijnands1999ApJ...526L..33W}
{Wijnands} R.,  {Homan} J.,   {van der Klis} M.,  1999, \mn@doi [\apjl]
  {10.1086/312365}, \href
  {https://ui.adsabs.harvard.edu/abs/1999ApJ...526L..33W} {526, L33}

\bibitem[\protect\citeauthoryear{{Xu} et~al.,}{{Xu}
  et~al.}{2018}]{Xu2018ApJ...852L..34X}
{Xu} Y.,  et~al., 2018, \mn@doi [\apjl] {10.3847/2041-8213/aaa4b2}, \href
  {https://ui.adsabs.harvard.edu/abs/2018ApJ...852L..34X} {852, L34}

\bibitem[\protect\citeauthoryear{{Yang}, {Brocksopp}, {Corbel}, {Paragi},
  {Tzioumis}  \& {Fender}}{{Yang} et~al.}{2010}]{Yang2010MNRAS.409L..64Y}
{Yang} J.,  {Brocksopp} C.,  {Corbel} S.,  {Paragi} Z.,  {Tzioumis} T.,
  {Fender} R.~P.,  2010, \mn@doi [\mnras] {10.1111/j.1745-3933.2010.00948.x},
  \href {https://ui.adsabs.harvard.edu/abs/2010MNRAS.409L..64Y} {409, L64}

\bibitem[\protect\citeauthoryear{Zhang et~al.,}{Zhang
  et~al.}{2022}]{Zhang_2022}
Zhang Y.,  et~al., 2022, \mn@doi [Monthly Notices of the Royal Astronomical
  Society] {10.1093/mnras/stac690}, 512, 2686

\bibitem[\protect\citeauthoryear{{Zhang} et~al.,}{{Zhang}
  et~al.}{2023}]{Zhang2023MNRAS.520.5144Z}
{Zhang} Y.,  et~al., 2023, \mn@doi [\mnras] {10.1093/mnras/stad460}, \href
  {https://ui.adsabs.harvard.edu/abs/2023MNRAS.520.5144Z} {520, 5144}

\makeatother
\end{thebibliography}



\bsp	
\label{lastpage}
\end{document}